\documentclass[prb,aps,twocolumn,showpacs,amsmath,amssymb,floatfix,superscriptaddress]{revtex4-1}

\usepackage{graphicx}
\usepackage{bm}
\usepackage{amsmath}
\usepackage{amssymb}
\usepackage{amssymb, verbatim}
\usepackage[normal]{subfigure}

\usepackage{hyperref}
\hypersetup{
        colorlinks=true,
}

\makeatletter

\usepackage{color}

\newcommand{\be}{\begin{equation} }
\newcommand{\ee}{\end{equation} }
\newcommand{\ba}{\begin{eqnarray} }
\newcommand{\ea}{\end{eqnarray} }
\newcommand{\ket}[1]{|#1\rangle}

\newcommand{\up}{\uparrow}
\newcommand{\dn}{\downarrow}
\renewcommand{\vec}[1]{\mathbf{ #1 }}
\DeclareMathOperator{\Tr}{Tr}
\DeclareMathOperator{\re}{Re}
\DeclareMathOperator{\im}{Im}
\DeclareMathOperator{\sgn}{sgn}

\begin{document}

\title{Signatures of Majorana Kramers Pairs in superconductor-Luttinger liquid and superconductor-quantum dot-normal lead junctions}

\author{Younghyun Kim}
\affiliation{Department of Physics, University of California, Santa Barbara, California 93106, USA}
\author{Dong E. Liu}
\affiliation{Station Q, Microsoft Research, Santa Barbara, California 93106-6105, USA}
\author{Erikas Gaidamauskas}
\affiliation{Center for Quantum Devices, Niels Bohr Institute, University of Copenhagen, DK-2100 Copenhagen, Denmark}
\author{Jens Paaske}
\affiliation{Center for Quantum Devices, Niels Bohr Institute, University of Copenhagen, DK-2100 Copenhagen, Denmark}
\author{Karsten Flensberg}
\affiliation{Center for Quantum Devices, Niels Bohr Institute, University of Copenhagen, DK-2100 Copenhagen, Denmark}
\author{Roman M. Lutchyn}
\affiliation{Station Q, Microsoft Research, Santa Barbara, California 93106-6105, USA}

\date{\today}

\begin{abstract}
Time-reversal invariant topological superconductors are characterized by the presence of Majorana Kramers pairs localized at defects. One of the transport signatures of Majorana Kramers pairs is the quantized differential conductance of $4e^2/h$ when such a one-dimensional  superconductor is coupled to a normal-metal lead. The resonant Andreev reflection, responsible for this phenomenon, can be understood as the boundary condition change for lead electrons at low energies.
In this paper, we study the stability of the Andreev reflection fixed point with respect to electron-electron interactions in the Luttinger liquid. We first calculate the phase diagram for the Luttinger liquid-Majorana Kramers pair junction and show that its low-energy properties are determined by Andreev reflection scattering processes in the spin-triplet channel, i.e. the corresponding Andreev boundary conditions are similar to that in a spin-triplet superconductor - normal lead junction. We also study here a quantum dot coupled to a normal lead and a Majorana Kramers pair and investigate the effect of local repulsive interactions leading to an interplay between Kondo and Majorana correlations. Using a combination of renormalization group analysis and slave-boson mean-field theory, we show that the system flows to a new fixed point which is controlled by the Majorana interaction rather than the Kondo coupling. This Majorana fixed point is characterized by correlations between the localized spin and the fermion parity of each spin sector of the topological superconductor. We investigate the stability of the Majorana phase with respect to Gaussian fluctuations.
\end{abstract}


\maketitle

\section{Introduction}

The search for topological superconductors, which host Majorana zero modes (MZMs),
has becomes an active pursuit in condensed matter physics \cite{Reich,Brouwer_Science,Wilczek2012,AliceaRev}.
Such exotic modes are predicted to obey non-Abelian braiding statistics\cite{Moore1991,Nayak1996,ReadGreen},
and have potential application in topological quantum computations \cite{kitaev,TQCreview}.
Many theoretical proposals for realizing topological superconductors in the laboratory have been
put forward recently~\cite{Fu&Kane08,Fu&Kane09,Sau10,Alicea10,LutchynPRL10,1DwiresOreg,MajoranaTInanowires,SauNature2012,Nadj-Perge13}, and, more excitingly, devices for detecting MZMs were successfully fabricated in the laboratory and the preliminary signatures of MZMs were
observed~\cite{Mourik2012,Das2012,Deng2012,Fink2012,Churchill2013,Nadj-Perge14,Deng2014,Higginbotham15,Albrecht16,HaoZhang16}.
Most research activity has focused on the topological superconductors belonging to class D ( {\it i.e.},
SCs with broken time-reversal-symmetry) and supporting an odd number of MZMs at a topological defect~\cite{Altland'97, TIClassification, Kitaev2009}.
However, Majorana zero modes can also appear in pairs in time-reversal invariant topological superconductors (TRITOPS)
belonging to class ${\rm DIII}$ \cite{ TIClassification, Kitaev2009, Teo&Kane10}. Those MZM pairs are referred to as
``Majorana Kramers pairs'' (MKPs), and their stability is protected by the time-reversal (TR) symmetry and the quasiparticle excitation gap. Recently, several theoretical proposals were put forward to realize TRITOPS~\cite{wong&Law12,dengPRL12,zhangPRL13,NakosaiPRL13,KeselmanPRL13,Gaidamauskas14,Klinovaja14,SchradePRL15}.
Transport signatures of MKPs and their detection schemes using a QPC were also recently investigated in a quantum spin Hall system~\cite{Li15}.

\begin{figure}[t]
\includegraphics[width=9cm]{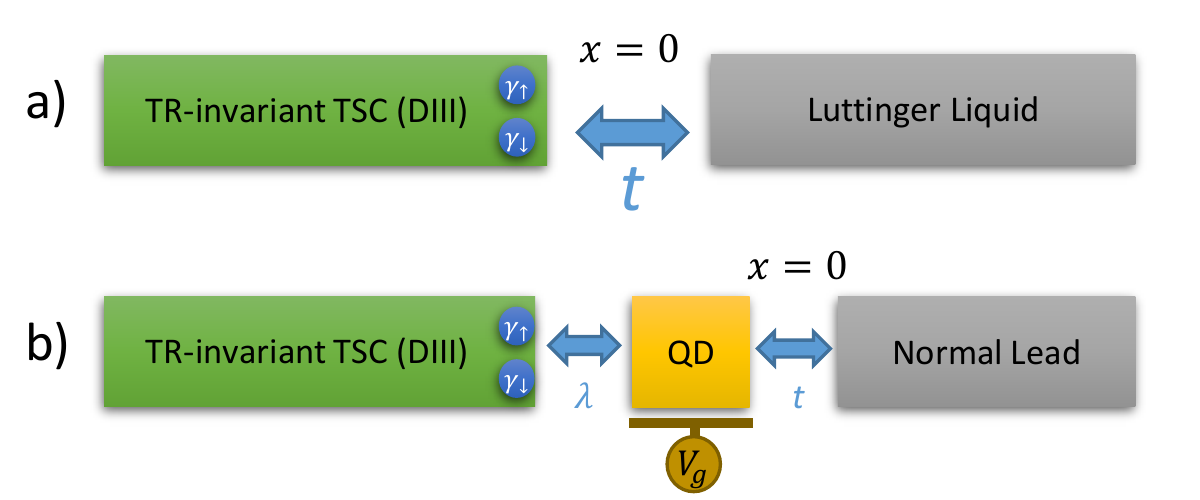}
\caption{(Color online) a) Schematic setup consisting of a) a junction between the Luttinger liquid and TRITOPS,
and b) a quantum dot coupled to a normal lead and a TRITOPS. Here, $x=0$ denotes the point in the lead which couples to the MKPs or quantum dots.}
\label{fig:device}
\end{figure}

Most previous works on MKPs considered non-interacting (or effectively non-interacting) models. It is well-known, however,
that interactions in one-dimensional systems are very important~\cite{Gangadharaiah11, Lobos'12, Fidkowski'12, Affleck'13} and
in some cases may even modify the classification of non-interacting systems~\cite{Fidkowski'11}.
For non-interacting systems, the presence of a MKP leads to a quantized conductance of $4e^2/h$ due to
perfect Andreev reflection at the junction. This quantization of the conductance is due to the  constraints imposed by TR symmetry which leads to complete decoupling of MKP in the non-interacting models. The situation is different, however, in the presence of interparticle interactions, and the fate of the perfect Andreev reflection fixed point is unclear. In this paper,
we study the  stability of MKPs with respect to electron-electron interactions and consider two generic systems - a) MKP
coupled to an interacting Luttinger liquid (see Fig.~\ref{fig:device} a)); b)  MKP coupled to an interacting quantum dot  (see Fig.~\ref{fig:device} b)).

We first consider a spinful Luttinger liquid lead with $\mathbb{SU}(2)$ spin symmetry coupled to a TRITOPS with a single MKP per end. In this case the boundary problem has an additional $\mathbb{U}(1)$ symmetry. We find that for weak repulsive interactions, $1>K_{\rho}\gtrsim 1/3$ with $K_\rho$ being the Luttinger parameter, the Andreev reflection fixed point ($\mathbb{A\times A}$) is stable
and the normal reflection fixed point ($\mathbb{N\times N}$) is unstable. For intermediate interaction strength $1/4<K_{\rho}\lesssim 1/3$,
the phase diagram depends on microscopic details, i.e. on the strength of four-fermion interactions allowed by TR symmetry,
which causes a Berezinsky-Kosterlitz-Thouless~(BKT)
transition between $\mathbb{A\times A}$ and $\mathbb{N\times N}$. Finally, for sufficiently strong repulsive interactions $K_{\rho}<1/4$,
the two electron backscattering term becomes relevant, and drives the system to a stable normal reflection fixed point.

In the presence of spin-orbit coupling, the corresponding boundary problem may break $\mathbb{U}(1)$ symmetry.  In this case, allowed processes involve spin-preserving and spin-flip Andreev scattering which drive the system to different boundary conditions for lead electrons: spin-preserving Andreev boundary ($\mathbb{A\times A}$) condition corresponds to $\psi_\sigma(0)=-\psi^\dag_\sigma(0)$ and spin-flip Andreev boundary ($\mathbb{SFA}$) condition corresponds to $\psi_\sigma(0)= -i\psi^\dag_{-\sigma}(0)$.
Thus, the corresponding phase diagram depends on
the relative strength of the corresponding Andreev scattering amplitudes. We find that the boundary conditions in this case are similar to those in a spin-triplet superconductor-Luttinger liquid junction and are stable with respect to weak repulsive interactions. In this sense, the physics is fundamentally different from an s-wave superconductor-Luttinger junction where weak repulsive interactions destabilize Andreev reflection fixed point~\cite{Fidkowski'12}.

In this paper, we also study the effect of local repulsive interactions by considering a MKP coupled to a quantum dot (QD) and
an $\mathbb{SU}(2)$-invariant normal lead (NL). In the limit of a large Coulomb interaction in the QD and single-electron occupancy, we investigate the competition between Kondo and Majorana correlations. When the coupling to the MKP is absent ($\lambda=0$), the system flows to the Kondo fixed point with the corresponding boundary conditions for NL electrons $\psi_{R\sigma}(0)=-\psi_{L\sigma}(0)$ where $R/L$ denote right and left movers. As we increase the coupling constant $\lambda$, the system exhibits a crossover
from the Kondo dominated regime to a Majorana dominated regime where the QD spin builds up a strong correlation with the MKP. The latter is characterized by $\mathbb{A\times A}$ boundary conditions $\psi_{R\sigma}(0)=-\psi^\dag_{L\sigma}(0)$. Thus, the problem at hand represents a new class of boundary impurity problems where spin in the dot is coupled to the fermion parity of a topological superconductor.

In order to understand thermodynamic and transport properties of this Majorana fixed point,
we have developed a slave-boson mean-field theory (please refer to Refs. \cite{Coleman84,BickersRMP87} for Anderson impurity models)
for this system. We show that the Majorana dominated regime corresponds to a new (i.e. different from Kondo) saddle-point solution.
We have analyzed the stability of this mean-field solution with respect to Gaussian fluctuations (in the spirit of Refs. \cite{Read&Newns83, Coleman'87}) finding that the mean field theory is stable (in the quasi-long range order sense) and can be used to calculate different observable quantities. We use this approach to calculate differential tunneling conductance as a function of applied voltage bias.

The paper is organized as follows. In Secs.~\ref{sec:SU2NoRashba} and \ref{sec:RashbaEffect}, we introduce the model of a MKP - Luttinger liquid junction, and consider the boundary problem with and without (e.g., due to Rashba spin-orbit coupling in the lead) $\mathbb{U}(1)$ symmetry. In Sec. \ref{sec:QD}, we study the signatures of a MKP in a QD-NL junction using both the renormalization group (RG) analysis
and the slave-boson mean-field theory. We also consider the Gaussian fluctuations around the mean-field solution, and analyze the stability
of the slave-boson mean-field solution. Finally, we conclude in Sec. \ref{sec:conclusion}.

\section{Majorana Kramers pair - Luttinger liquid junction}\label{sec:MKPLL}

In this section we consider the setup shown in Fig. \ref{fig:device} a) consisting of a semi-infinite spinful Luttinger liquid coupled weakly to a TRITOPS. We assume that the topological gap of the superconductor is much larger than the other relevant energy scales
(i.e. tunneling amplitudes $t_\sigma$, $\Delta$ and $\Delta_{\rm AN}$ see the text below Eq. \eqref{eq:model0} for definitions). Thus, in the low-energy approximation the superconductor
Hamiltonian consists of only the MKPs localized at its opposite ends. In this section, we will use $\psi_{\sigma}(0)$ to describe the operators at the boundary
$x=0$, and use $t(l_0)$ (similarly for $\tilde{t}$, $\Delta$ and $\tilde{\Delta}$) as the initial value in RG flow with the initial length cutoff $l_0$.

\subsection{Majorana Kramers pair coupled to $\mathbb{SU}(2)$-invariant Luttinger liquid}\label{sec:SU2NoRashba}

\subsubsection{Theoretical Model}

We first consider an $\mathbb{SU}(2)$-invariant interacting nanowire coupled to a MKP. The Hamiltonian for the 1D lead can be written as the spinful Luttinger model
\begin{equation}
 H_{\rm{lead}} = \sum_{j=\rho,\sigma}\frac{v_{j}}{2\pi}\int_{0}^{\infty}dx\left(K_{j}(\partial_{x}\theta_{j})^{2}+\frac{(\partial_{x}\phi_{j})^{2}}{K_{j}}\right)
\label{eq:Hlead}
\end{equation}
where  $v_{\rho/\sigma}$ and $K_{\rho/\sigma}$are velocity and
Luttinger parameter for charge and spin modes, respectively.
The bosonic fields satisfy the commutation relation
$[\phi_{\alpha}(x),\theta_{\beta}(x')]=i \pi K_{\alpha} \delta_{\alpha\beta}\text{sign}(x-x')$. We use here the following convention for the Abelian bosonization procedure~\cite{giamarchi}:
\begin{equation}
\psi_{R/L,s}(x)= \frac{\Gamma_{R/L,s}}{\sqrt{2\pi a}}e^{i\frac{1}{\sqrt{2}}\left\{ \pm[\phi_{\rho}(x)+s\phi_{\sigma}(x)]+\theta_{\rho}(x)+s\theta_{\sigma}(x)\right\} }
\end{equation}
where $R/L$ represents right/left moving modes, $a$ is an ultraviolet cutoff length scale, $s=\uparrow/\downarrow$
denotes fermion spin, and $\Gamma_{R/L,s}$ is the Klein factor.

The total Hamiltonian is given as
\begin{align}\label{eq:totalH}
H=H_{\rm lead}+H_B.
\end{align}
where $H_B$ the coupling between the Luttinger liquid lead and the MKP. We neglect here the ground-state degeneracy splitting energy.  The most general form of the TR invariant boundary Hamiltonian describing the coupling between the MKP and Luttinger liquid and including only two and four-fermion operators reads
\begin{align}
H_B &=i\, t_{\uparrow}\gamma_{\uparrow}\left(\psi_{\uparrow}(0)+\psi_{\uparrow}^{\dagger}(0)\right)
         -i\, t_{\downarrow}\gamma_{\downarrow}\left(\psi_{\downarrow}(0)+\psi_{\downarrow}^{\dagger}(0)\right)  \nonumber\\
 & -\Delta i\gamma_{\uparrow}\gamma_{\downarrow}\left(-i\psi_{\uparrow}^{\dagger}(0)\psi_{\downarrow}(0)+i\psi_{\downarrow}^{\dagger}(0)\psi_{\uparrow}(0)\right)  \nonumber\\
 &-\Delta_{\rm AN} i\gamma_{\uparrow}\gamma_{\downarrow}\left(-i\psi_{\uparrow}^{\dagger}(0)\psi_{\downarrow}^{\dagger}(0)+i\psi_{\downarrow}(0)\psi_{\uparrow}(0)\right)
\label{eq:model0}
\end{align}
where $\gamma_{\up/\dn}$ are Majorana operators with $\{\gamma_\sigma,\gamma_{\sigma'}\}=2\delta_{\sigma,\sigma'}$. Here $t_{\up/\dn}$, $\Delta$
and $\Delta_{\text{AN}}$ are set to be real. The first two terms represent tunneling between the lead and the MKP with the
amplitudes $t_{\uparrow/\downarrow}$. Under TR symmetry $T$, field operators $\psi$ transform as
\begin{eqnarray}
 &T \psi_{R/L\uparrow} = \psi_{L/R\downarrow},\\
 &T \psi_{R/L\downarrow} = -\psi_{L/R\uparrow},
\end{eqnarray}
(i.e. $T^2=-1$) and the coupling constants in the Hamiltonian need to complex conjugated. TR symmetry requires that $t_{\uparrow}=t_{\downarrow}=t$ with $t$ being real.
Assuming the spin-quantization axis is fixed in the whole system, the overall Hamiltonian $H$
has $\mathbb{U}(1)$ spin-rotation symmetry, leaving it invariant under the unitary transformation:
\begin{align}
 (\psi_{\uparrow}, \psi_{\downarrow})^T &\rightarrow R(\theta)  (\psi_{\uparrow}, \psi_{\downarrow})^T \\
 (\gamma_{\uparrow},\gamma_{\downarrow})^T & \rightarrow R(-\theta)   (\gamma_{\uparrow},\gamma_{\downarrow})^T.
\label{eq:U1symmetry}
\end{align}
Here $R(\theta)=e^{i\frac{\theta}{2}\sigma_y}$ represents a $\mathbb{U}(1)$ spin-rotation matrix by an angle $\theta$.  Thus, electron tunneling between Luttinger liquid and topological superconductor preserves the spin.
The last two terms $\Delta$ and $\Delta_{\rm AN}$
represent normal, and anomalous backscattering terms,
which, in fact, will also be generated by the tunneling terms in the RG flow in the presence of interactions in the Luttinger liquid.

\subsubsection{Weak coupling RG analysis near normal reflection fixed point}

We now study the stability of the weak coupling normal reflection fixed point using perturbative RG analysis.
In the ultraviolet, the boundary conditions for lead electrons at $x=0$ are given by $\psi_{R\sigma}(0)=\psi_{L\sigma}(0)$ (i.e. perfect normal reflection). In terms of bosonization language, this boundary condition corresponds to $\Gamma_{L,s}=\Gamma_{R,s}$ and pinning $\phi_{\rho,\sigma}(0)$.
Once we turn on the boundary couplings $t$, $\Delta$ and $\Delta_{\rm AN}$, boundary conditions for lead electrons may change depending on the strength of interaction in the lead.
Let us study now the stability of this normal reflection fixed point. After integrating out the fields away from $x=0$, the corresponding imaginary-time partition function becomes
\begin{equation}
\mathcal{Z}={\displaystyle \int}D[\theta_{\rho}]D[\theta_{\sigma}]\; e^{-(S_{0}+S_{T})},
\end{equation}
with
\begin{equation}
S_{0}=\sum_{j=\rho,\sigma}\frac{K_{j}}{2\pi}\int\frac{d\omega}{2\pi}|\omega||\theta_{j}(\omega)|^{2},
\label{eq:ActionBC_Normal}
\end{equation}
and the boundary coupling term reads
\begin{align}
S_{T}&=\int \frac{d\tau}{2\pi a}\Bigg[t\big( i\gamma_{\uparrow}\Gamma_{\uparrow}\cos\frac{\theta_{\rho}+
\theta_{\sigma}}{\sqrt{2}}-i\gamma_{\downarrow}\Gamma_{\downarrow}\cos\frac{\theta_{\rho}-\theta_{\sigma}}{\sqrt{2}}\big)\nonumber\\
 & -\Delta\gamma_{\uparrow}\gamma_{\downarrow}\Gamma_{\uparrow}\Gamma_{\downarrow}\cos\sqrt{2}\theta_{\sigma}
 -\Delta_{\rm AN}\gamma_{\uparrow}\gamma_{\downarrow}\Gamma_{\uparrow}\Gamma_{\downarrow}\cos\sqrt{2}\theta_{\rho}\Bigg],
\end{align}
where $a$ is the ultraviolet cutoff. Here we used short-hand notation $\theta_{j}(\tau)$ denoting the fields at $x=0$.

We now perform a perturbative RG procedure by separating the bosonic fields $\theta_{j}$ into slow, and fast modes and integrating out the fast modes. After some manipulations, the new effective action can be calculated using the
cumulant expansion:
\begin{equation}
S_{{\rm eff}}[\theta_{j}^{<}]=S_{{\rm 0}}[\theta_{j}^{<}]+\langle S_{T}\rangle-\frac{1}{2}\left(\langle S_{T}^{2}\rangle-\langle S_{T}\rangle^{2}\right),
\end{equation}
where the average $\langle\cdots\rangle$ describes an integration
over the fast modes.  The details of this calculation are presented in the Appendix \ref{app:RG2order_SU2}, and we simply summarize the RG equations here
\begin{eqnarray}
\frac{dt}{dl} & = & \left(1-\frac{1}{4K_{\rho}}-\frac{1}{4K_{\sigma}}\right)t -\frac{\Delta t }{4\pi v K_{\sigma}} -\frac{\Delta_{\rm AN} t}{4\pi v K_{\rho}},\,\,\,\,\,\,\,\,\label{eq:U1RG_t} \\
\frac{d\Delta}{dl} & = & \left(1-\frac{1}{K_{\sigma}}\right)\Delta-\left(\frac{1}{K_{\rho}}-\frac{1}{K_{\sigma}}\right)\frac{t^2}{4\pi v} , \label{eq:U1RG_Delta} \\
\frac{d\Delta_{\rm AN}}{dl} & = & \left(1-\frac{1}{K_{\rho}}\right)\Delta_{\rm AN}+\left(\frac{1}{K_{\rho}}-\frac{1}{K_{\sigma}}\right)\frac{t^2}{4\pi v}. \label{eq:U1RG_delta}
\end{eqnarray}
Here $dl=d\ln b$ where $b$ is the ratio of the cutoff change from $\Lambda$ to $\Lambda/b$ with $\Lambda =v/a$.
One can notice that $t$ is a relevant perturbation and grows under RG. Therefore, in the non-interacting case when $\Delta, \Delta_{\rm AN}=0$,
the system will flow to the perfect Andreev reflection fixed point ($\mathbb{A\times A}$) corresponding to the boundary condition $\psi_{L,s}^{\dagger}(0)=-\psi_{R,s}(0)$~\cite{Fidkowski'12} and quantized differential conductance $G=\frac{4e^2}{h}$ at zero temperature.

\begin{figure}
\includegraphics[scale=0.55]{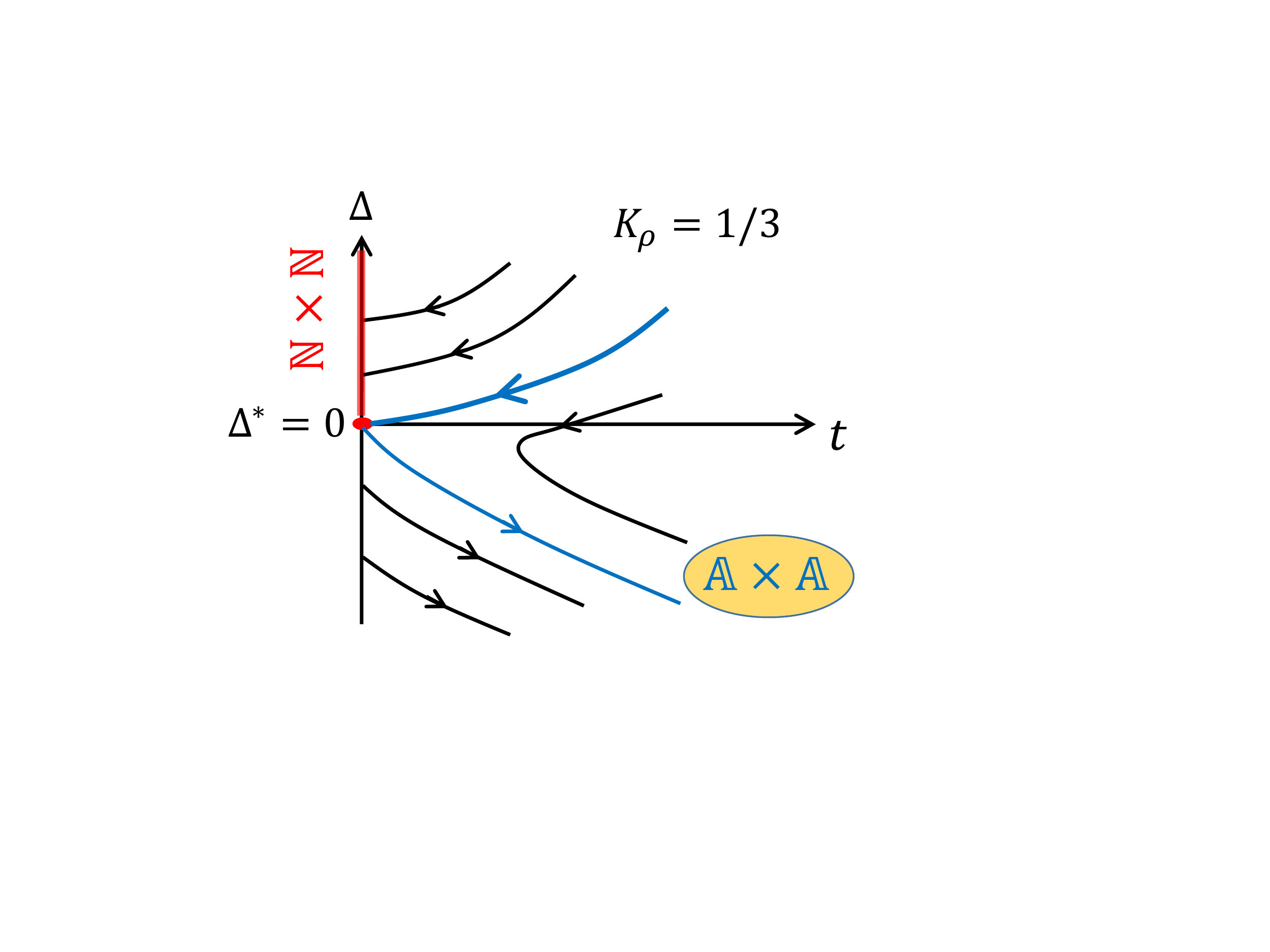}
\caption{Illustration of the RG flow diagram near the normal reflection fixed point $\mathbb{N\times N}$ for the case with $U(1)$ spin symmetry.
The red line indicates the regime where $\mathbb{N\times N}$ is stable. Here, we set $K_\rho=1/3$}
\label{fig:SU2symRGD}
\end{figure}

Let us now try to understand the effects of interactions. In this section, we will focus on an $\mathbb{SU}(2)$ spin-invariant lead ($K_{\sigma}=1$) and repulsive interactions in the nanowire $K_{\rho}<1$. In this case,
the coupling $\Delta_{\rm AN}$ is irrelevant and can be neglected, and RG equations simplify to
\begin{eqnarray}
\frac{dt}{dl} & = & \left(\frac{3}{4}-\frac{1}{4K_{\rho}}\right)t-\frac{\Delta  t}{4\pi v} ,\\
\frac{d\Delta}{dl} & = & - \left(\frac{1}{K_{\rho}}-1\right) \frac{ t^2}{4\pi v}.
\end{eqnarray}
The coupling $t$ is relevant for not too strong repulsive interactions. It becomes marginal, however, if initial value of $\Delta(l_0)$ is equal to the special value  $\Delta^* = \pi v (3-\frac{1}{K_\rho})$. Indeed, then above RG equations (after a slight redefinition of variables) are identical to the anisotropic Kondo model~\cite{giamarchi}, the solution of which is well-known. If the initial value of $\Delta (l_0)$ is zero, and the parameter $\Delta^*>0$ (i.e. $K_{\rho}>1/3$), the system will flow to strong coupling $\mathbb{A\times A}$ fixed point
whereas for $K_{\rho}\lesssim 1/3$, the system will flow to $\mathbb{N\times N}$ fixed point for small $t(l_0)$
and flow to strong coupling $\mathbb{A\times A}$ for larger $t(l_0)$. The perturbative RG flow is summarized in Fig. \ref{fig:SU2symRGD}.

\subsubsection{Weak coupling RG analysis near perfect Andreev reflection fixed point}

\begin{figure}
\includegraphics[scale=0.45]{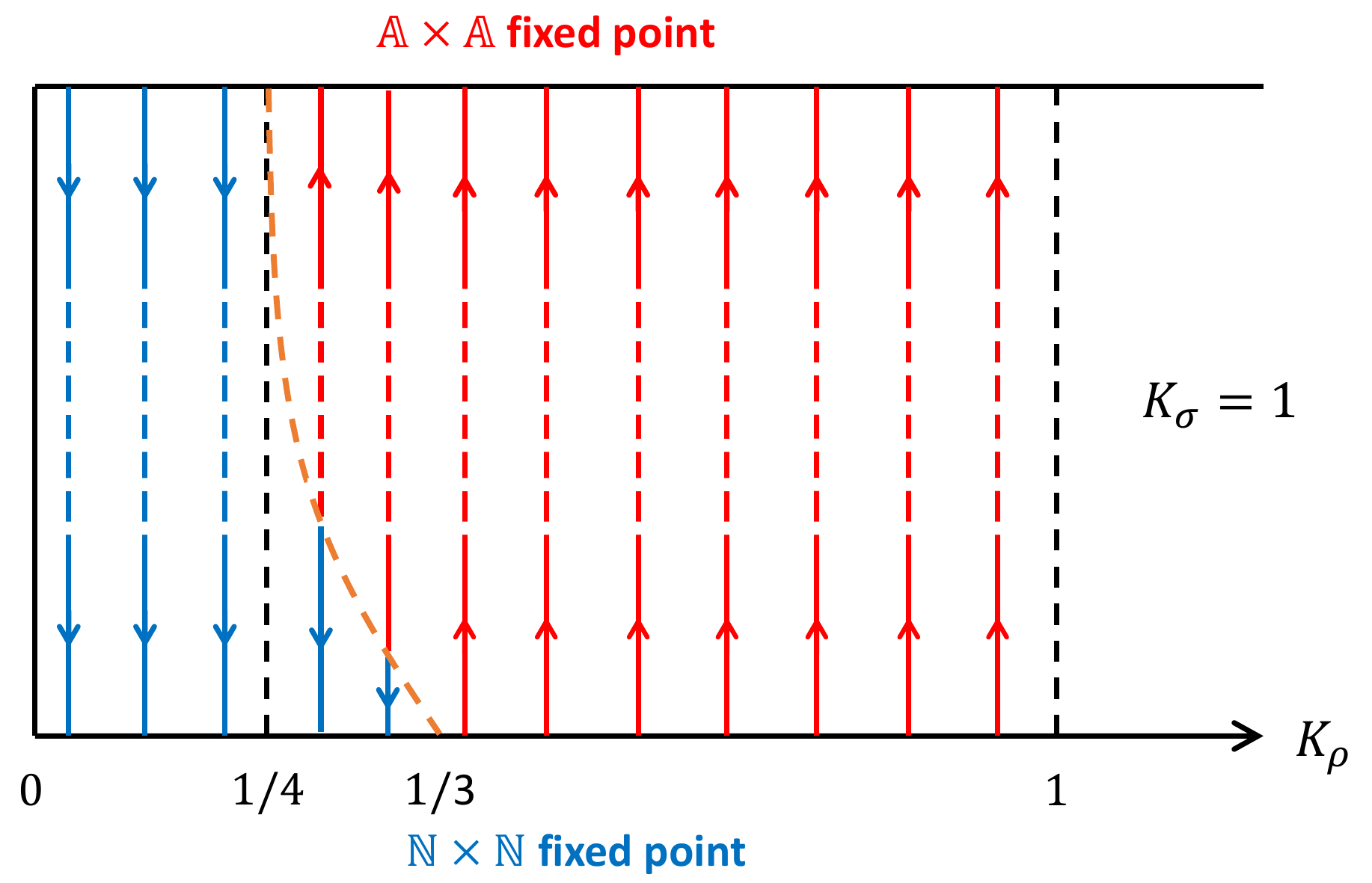}
\caption{Illustration of the flow between the normal reflection fixed point $\mathbb{N\times N}$
and the Andreev reflection fixed point $\mathbb{A\times A}$ for the case with $U(1)$ spin symmetry.
For the analysis at $\mathbb{N\times N}$, the boundary perturbation $\Delta$ bends the transition
line, i.e. the brown dashed line connecting $K_{\rho}=1/3$ at  $\mathbb{N\times N}$ and $K_{\rho}=1/4$ at  $\mathbb{A\times A}$.
We also assume the initial value $\Delta(l_0)$ in RG is zero.}
\label{fig:SU2symFlowD}
\end{figure}

As shown in the previous section, the normal reflection fixed point is unstable for weak repulsive interactions and the system flows to the perfect Andreev fixed point corresponding to the boundary conditions $\psi_{L,s}^{\dagger}(0)=-\psi_{R,s}(0)$ which, in bosonic variables corresponds to pinning $\theta_{\rho}$ and $\theta_{\sigma}$ fields at $x=0$. Thus, the fluctuating degrees of freedom are the fields $\phi_{\rho}$ and $\phi_{\sigma}$  and the corresponding boundary action reads
\begin{equation}
S_{0}=\sum_{j=\rho,\sigma}\frac{1}{2\pi K_{j}}\int\frac{d\omega}{2\pi}|\omega||\phi_{j}(\omega)|^{2}.
\label{eq:ActionBC_Andreev}
\end{equation}
We now consider perturbations near the Andreev fixed point which are consistent with time-reversal and the spin-$\mathbb{SU}(2)$ symmetry of the Luttinger liquid lead. The only fermion bilinear boundary perturbation preserving aforementioned  symmetries is
\begin{align}
 H_{1B}&=\lambda_1 (\psi_{R\uparrow}^{\dagger}(0)\psi_{L\uparrow}(0)+\psi_{R\downarrow}^{\dagger}(0)\psi_{L\downarrow}(0))+h.c.\nonumber\\
 &=\frac{\lambda_1}{2\pi a}\cos\left(\sqrt{2}\phi_{\rho}\right)\cos\left(\sqrt{2}\phi_{\sigma}\right).
 \label{eq:NormalBackscattering}
\end{align}
In addition, one has to also consider the following four-fermion perturbation consistent with the above symmetries:
\begin{align}
 H_{2B}=&\lambda_2 \psi_{L\uparrow}^{\dagger}(0)\psi_{R\uparrow}(0)\psi_{L\downarrow}^{\dagger}(0)\psi_{R\downarrow}(0)+h.c. \nonumber \\
&= \frac{\lambda_2}{(2\pi a)^2}\sin(2\sqrt{2}\phi_{\rho}),
 \label{eq:TwoBackscattering}
\end{align}
which corresponds to two-electron backscattering. The perturbative RG equations for $\lambda_1$ and $\lambda_2$ are given by
\begin{align}
\frac{d\lambda_1}{dl}&=(1-K_{\rho}-K_{\sigma})\lambda_1\\
\frac{d\lambda_2}{dl}&=(1-4K_{\rho})\lambda_2
\end{align}
One can see that the first term $\lambda_1$ is irrelevant since $K_\sigma=1$ whereas the second coupling becomes relevant for $K_{\rho}<1/4$ indicating that $\mathbb{A\times A}$ fixed point becomes unstable for strong repulsive interactions.
Taking into account the perturbative RG analysis near both $\mathbb{N\times N}$ and $\mathbb{A\times A}$ fixed points,
we conjecture the qualitative phase diagrams shown in Fig. \ref{fig:SU2symFlowD}.

\subsubsection{Differential tunneling conductance}

We now discuss transport signatures of MKPs. The simplest experiment to detect the presence of a MKP is the differential conductance measurement.
We focus on the case with an $SU(2)$ symmetric wire $K_{\sigma}=1$ and calculate $G=dI/dV$ at zero voltage bias as a function of temperature. The RG flow between the normal $\mathbb{N\times N}$ and Andreev reflection $\mathbb{A\times A}$ defines a crossover temperature $T^*$, which roughly corresponds to the width of the zero bias peak.
Although the conductance for the whole crossover regime requires involved calculations, the conductance around $\mathbb{N\times N}$
and $\mathbb{A\times A}$ fixed points can be obtained using perturbation theory, see, e.g., Ref.~[\onlinecite{Lutchyn2013}].

First of all, we consider the case $1/3<K_{\rho}<1$, where $\mathbb{A\times A}$ fixed point is stable.
In the ultraviolet (i.e. near the unstable normal reflection fixed point), the leading relevant perturbation is the coupling to the MKP, $t$, which has scaling dimension $\frac{3}{4}-\frac{1}{4K_{\rho}}$. Near the  stable Andreev reflection fixed point (i.e. in the infrared),
the deviation from the quantized value comes from the leading irrelevant operators which cause backscatterings,
i.e. single-electron backscattering shown in Eq. (\ref{eq:NormalBackscattering}) with scaling dimension $-K_{\rho}$ and
two-electron backscattering  shown in Eq. (\ref{eq:TwoBackscattering}) with scaling dimension $1-4K_{\rho}$.

Here, for $K_{\rho}>1/3$, the single-electron backscattering shown in Eq. (\ref{eq:NormalBackscattering}) is the leading irrelevant operator. We can now obtain scaling of the conductance with temperature at zero bias (assuming the initial value of $\Delta$ coupling is zero, i.e. $\Delta(l_0)=0$):
\begin{equation}
 \frac{G}{4e^2/h} \Bigg |_{K_{\rho}>\frac{1}{3}} = \begin{cases}
                        c_{1,T}(K_\rho)\left(\frac{T}{T^*}\right)^{2(\frac{1}{4K_{\rho}}-\frac{3}{4})}, & \quad T\gg T^* \\
                        1 - c_{2,T} (K_{\rho}) \left(\frac{T}{T^*}\right)^{2K_{\rho}}, & \quad T\ll T^* \\
                       \end{cases},
\end{equation}
where $c_{1/2,T}(K_{\rho})$ are numerical coefficients of the order one. Similarly, one can obtain voltage corrections to the conductance at zero temperature. Interestingly, the analogous coefficient $c_{1,V}(K_{\rho})$ vanishes in the non-interacting limit and, therefore, the scaling of the conductance with voltage and temperature is different at $K_{\rho}=1$, see Ref.~[\onlinecite{Lutchyn2013}] for details.

Next, we consider $K_{\rho}<1/4$, where $\mathbb{N\times N}$ is stable in the infrared.
In this case, we start near the high energy unstable fixed point $\mathbb{A\times A}$ and calculate the conductance by perturbing with the two-electron backscattering operator which is the leading relevant operator in this regime.
Thus, we obtain
\begin{equation}
 \frac{G}{4e^2/h} \Bigg |_{K_{\rho}\lesssim\frac{1}{3}} \sim \begin{cases}
                        1- c_{3,T}(K_{\rho}) \left(\frac{T}{T^*}\right)^{2(4K_{\rho}-1)}, & \quad T\gg T^* \\
                        c_{4,T}(K_\rho)\left(\frac{T}{T^*}\right)^{2(\frac{1}{4K_{\rho}}-\frac{3}{4})}, & \quad T\ll T^* \\
                       \end{cases},
\end{equation}
where $c_{3/4,T}(K_{\rho})$ are $\mathcal{O}(1)$ numerical coefficients.

The calculation of the conductance in the regime $1/4<K_{\rho}\lesssim 1/3$ depends on microscopic details (i.e. strength of $t(l_0)$), and is outside the scope of the paper.

\subsection{The effect of the Rashba spin-orbit coupling in the lead}\label{sec:RashbaEffect}

\subsubsection{Theoretical Model}\label{sec:IIB1}
In this section, we consider the effect of Rashba spin-orbit coupling (SOC) in the nanowire. When coupling to MKP, the spin eigenstates of the MKP do not have to be the same as the
the spin eigenstates of the nanowire. Therefore, tunneling between the lead and the TRITOPS will have both spin-preserving  and spin-flip components. In order to see how the spin flip tunneling
is generated, we consider the direction of the Rashba coupling which has an angle $\theta$ rotation
compared to that of the MKP. The corresponding tight binding model can written as
\begin{align}
 H &= H_{\rm lead} + H_{\rm T}\\
H_{\rm lead} &= -t \sum_{j=1}^{N}\sum_{s} \left(c_{j+1, s}^{\dagger}c_{j, s} +h.c.  \right)
       + \mu \sum_{j s} c_{j, s}^{\dagger}c_{j, s} \nonumber\\
     &  +\sum_{js s'} (-i) \alpha_R c_{j+1,s}^{\dagger} \left( \cos\theta \sigma_z +\sin\theta \sigma_y \right)_{ss'} c_{j,s'} +h.c.,\nonumber\\
 H_{\rm T}&= i t_0 \left[ \gamma_{\uparrow} ( c_{N\uparrow}+c_{N\uparrow}^{\dagger} )
 -\gamma_{\downarrow} ( c_{N\downarrow}+c_{N\downarrow}^{\dagger} )\right].
\end{align}
One can see that the above Hamiltonian respects TR symmetry. We apply the following unitary transformation
\begin{equation}
 \left( \begin{array}{c}
         d_{i\uparrow} \\
         d_{i\downarrow}
        \end{array}
 \right) = e^{-i\frac{\theta}{2}\sigma_x}
  \left( \begin{array}{c}
         c_{i\uparrow} \\
         c_{i\downarrow}
        \end{array}
 \right),
\end{equation}
and then the bulk and boundary Hamiltonians become
\begin{eqnarray}
 H_{\rm lead} &=& \mu \sum_{js}d_{j, s}^{\dagger}d_{j, s} + \sum_j \Big[ (-t-i\alpha_R)d_{j+1, \uparrow}^{\dagger}d_{j, \uparrow} \nonumber\\
      &&  + (-t+i\alpha_R)d_{j+1, \downarrow}^{\dagger}d_{j, \downarrow} +h.c.\Big] ,
\end{eqnarray}
and
\begin{eqnarray}
 H_{\rm T} &= & i t \sum_{s=\uparrow,\downarrow} s \gamma_{s} (d_{N,s}+d_{N,s}^{\dagger})\nonumber\\
         & & + \tilde{t} \sum_{s} s \gamma_{s} (d_{N,-s}^{\dagger}-d_{N,-s}),
\end{eqnarray}
where $t=t_0 \cos\theta$ and $\tilde{t}=t_0 \sin\theta$, and $s=1(-1)$ for spin-$\up(\dn)$.
Therefore, the spin-flip tunneling is non-zero for any $\theta\neq 0$, i.e. due to the presence of SOC.
One can simply check that, in the presence of both $t$ and $\tilde t$, the $U(1)$ symmetry shown in Eq. (\ref{eq:U1symmetry}) is broken.
In this case, the boundary condition at the Andreev reflection fixed point is determined by the relative magnitude of $t$ and $\tilde t$. For the discussion of boundary condition and bosonization procedure in the normal reflection fixed point, please refer to Appendix \ref{app:soc_boundary}.

It is instructive to analyze the boundary conditions in the non-interacting case using the scattering matrix approach. The unitary scattering matrix
is defined as (see, e.g., Ref.~\cite{Nilsson08})
\begin{equation}\label{eq:Sw}
 S(\omega) = \hat{I} + 2\pi i \hat{W}^{\dagger} \left( H_{MK} -\omega - i\pi \hat{W}\hat{W}^{\dagger} \right)^{-1} \hat{W},
\end{equation}
where $H_{MK}$ is the Hamiltonian for the MKP (2 by 2 matrix) which vanishes in the limit $L \gg \xi$ with $L$ and $\xi$ being respectively the length and coherence length of the superconductor. Note that the local term $i \delta E \gamma_{\uparrow} \gamma_{\downarrow}$ is not allowed by TR symmetry. The matrix $\hat{W}$
describes the coupling between the MKP ${\gamma_{\uparrow},\gamma_{\downarrow}}$ and the lead degrees of
freedom in the basis $({\psi_{\uparrow},\psi_{\downarrow},\psi_{\uparrow}^{\dagger},\psi_{\downarrow}^{\dagger}})$:
\begin{equation}
 \hat{W} =
 \begin{pmatrix}
  it & \tilde{t} & it & -\tilde{t} \\
  -\tilde{t} & -it & \tilde{t} & -it
 \end{pmatrix}.
\end{equation}
Note that we assume that lead Hamiltonian is diagonal here. Therefore, $\psi_\sigma$ represent helicity eigenstates in the case of a Rashba model. Using Eq. \eqref{eq:Sw}, we can represent the scattering matrix at $\omega=0$ as
\begin{equation}
 S(0)=
 \begin{pmatrix}
  S^{ee}(0) & S^{eh}(0) \\
  S^{he}(0) & S^{hh}(0)
 \end{pmatrix}.
\end{equation}
The components $S^{ee}(0)$ and $S^{eh}(0)$ describe normal and Andreev reflection, respectively. As pointed out in Ref. \cite{Li15}, the normal part $S^{ee}(0)$ is zero so we focus on the non-diagonal components:
\begin{align}
 S^{eh}(0) &=
 \begin{pmatrix}
    \frac{\tilde{t}^2-t^2}{t^2+\tilde{t}^2} &  -\frac{2i \tilde{t} t}{t^2+\tilde{t}^2} \\
    -\frac{2i \tilde{t} t}{t^2+\tilde{t}^2}  &  \frac{\tilde{t}^2-t^2}{t^2+\tilde{t}^2}
 \end{pmatrix},\nonumber\\
&=-\cos 2\theta-i\sigma_x\sin 2\theta
\end{align}
where the diagonal term is the coefficient of the same-spin Andreev reflection $\psi_{\uparrow}\rightarrow \psi_{\uparrow}^{\dagger}$,
and the off-diagonal term is the coefficient of the spin-flip Andreev reflection $\psi_{\uparrow}\rightarrow\psi_{\downarrow}^{\dagger}$.
As we change the angle of SOC, $\theta$, from $0\,(\tilde{t}=0)$ to $\pi/4\,(t=\tilde{t})$, the Andreev reflection boundary condition
changes continuously from $\psi_{L,s}(0)=-\psi^\dagger_{R,s}(0)(\mathbb{A\times A})$ with $s=\uparrow,\downarrow$
(i.e. $t\neq 0$ and $\tilde{t}=0$)
to $\psi_{L\up}(0)=-i\psi^\dagger_{R\dn}(0)$ and $\psi_{L\dn}(0)=-i\psi^\dagger_{R\up}(0)$.
We denote this boundary condition for $t=\tilde{t}$ as \textit{spin flip Andreev reflection boundary condition} $(\mathbb{SFA})$,
which describes an Andreev reflection with spin-flip processes.
Upon increasing $\theta$ to $\pi/2$, the boundary condition becomes $\psi_{L,s}(0)=\psi^\dagger_{R,s}(0) (\mathbb{\tilde{A}\times \tilde{A}})$
(i.e. $t= 0$ and $\tilde{t}\neq 0$).
\begin{figure}
	\centering
  \includegraphics[scale=0.50]{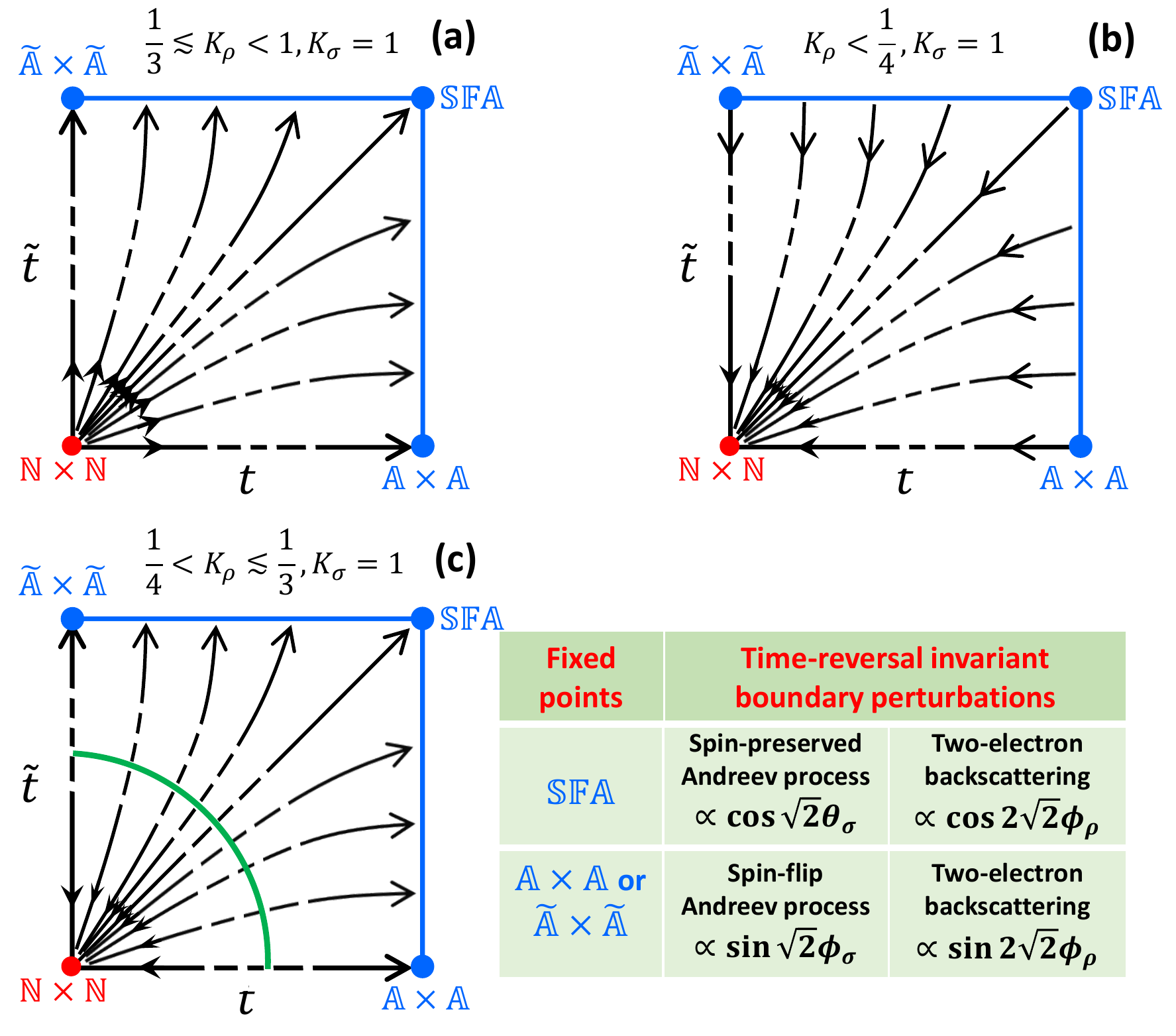}
\caption{RG flow diagram for the junction without $U(1)$ symmetry: (a) for $1/3<K_{\rho}<1$ and $K_{\sigma}=1$,
(b) for $K_{\rho}<1/4$ and $K_{\sigma}=1$, and (c) for $1/4<K_{\rho}\lesssim 1/3$ and $K_{\sigma}=1$, and the green line indicates
the conjectured BKT phase transition. The inset table summarizes
the important time-reversal invariant boundary perturbations near $\mathbb{SFA}$, $\mathbb{A\times A}$, and $\mathbb{\tilde{A}\times \tilde{A}}$ fixed points.
For $K_{\sigma}=1$, along each line of the RG flow, the phase diagram as a function of $K_{\rho}$ is similar to the that shown in Fig.~\ref{fig:SU2symRGD}.}
\label{fig:RGflow}
\end{figure}
Here we would like to emphasize that the $\mathbb{SFA}$ boundary condition is different from the Andreev boundary condition in s-wave
spin-singlet superconducting junction where $\psi_{L\up}(0)=\mp i\psi^\dagger_{R\dn}(0)$ and $\psi_{L\dn}(0)=\pm i\psi^\dagger_{R\up}(0)$ (see, e.g., Ref.\onlinecite{maslovPRB96}). Notice different signs in this case for spin-up and spin-down components. The $\mathbb{SFA}$ boundary condition in our case corresponds to spin-triplet Andreev reflection which typically is realized at junctions between a normal lead and a spin-triplet p-wave superconductor.  Indeed, if we denote spin-triplet  pair potential as $\Delta(p) \propto (\overrightarrow{d}(p)\cdot \overrightarrow{\sigma})i\sigma_{y}$, then different orientations of the $\overrightarrow{d}$-vector correspond to  $\mathbb{SFA}$ ($\overrightarrow{d}\propto (0,0,1)$) and $\mathbb{A \times A}$ ($\overrightarrow{d} \propto (0,\pm 1,0)$) boundary conditions. This difference between conventional (s-wave) spin-singlet Andreev boundary conditions and $\mathbb{SFA}$ boundary conditions considered here becomes very important later when we consider allowed boundary perturbations.

\subsubsection{RG analysis near normal reflection fixed point $\mathbb{N\times N}$}

Let's now analyze the interaction effects in the lead. In the absence of U(1) spin-rotation symmetry, we can have additional terms in the boundary action:
\begin{eqnarray}
S_{\rm T} & = & \int d\tau \Bigg[ i\, t\gamma_{\uparrow}\left(\psi_{\uparrow}(0)+\psi_{\uparrow}^{\dagger}(0)\right)
                 -i\, t\gamma_{\downarrow}\left(\psi_{\downarrow}(0)+\psi_{\downarrow}^{\dagger}(0)\right) \nonumber\\
  & & +\tilde{t}\gamma_{\uparrow}\left(\psi_{\downarrow}(0)-\psi_{\downarrow}^{\dagger}(0)\right)-\tilde{t}\gamma_{\downarrow}\left(\psi_{\uparrow}(0)-\psi_{\uparrow}^{\dagger}(0)\right)\nonumber\\
 &  & -\Delta i\gamma_{\uparrow}\gamma_{\downarrow}\left(-i\psi_{\uparrow}^{\dagger}(0)\psi_{\downarrow}(0)+i\psi_{\downarrow}^{\dagger}(0)\psi_{\uparrow}(0)\right) \nonumber\\
  & & +\tilde{\Delta}i\gamma_{\uparrow}\gamma_{\downarrow}\left(\psi_{\uparrow}^{\dagger}(0)\psi_{\uparrow}(0)-\psi_{\downarrow}^{\dagger}(0)\psi_{\downarrow}(0)\right) \Big].
\end{eqnarray}
We have omitted here the irrelevant terms, e.g. $\Delta_{\text{AN}}$, analogous to those considered in Sec. \ref{sec:SU2NoRashba}.
After the bosonization, the boundary action reads
\begin{eqnarray}
 S_{T}&=&\int d\tau \Bigg[ \frac{t}{2\pi a} \left( i\gamma_{\uparrow}\Gamma_{\uparrow}\cos\frac{\theta_{\rho}+
         \theta_{\sigma}}{\sqrt{2}}-i\gamma_{\downarrow}\Gamma_{\downarrow}\cos\frac{\theta_{\rho}-\theta_{\sigma}}{\sqrt{2}}\right) \nonumber\\
     &&    + \frac{\tilde{t}}{2\pi a}\left(i\gamma_{\downarrow}\Gamma_{\uparrow}\sin\frac{\theta_{\rho}+\theta_{\sigma}}{\sqrt{2}}
         - i\gamma_{\uparrow}\Gamma_{\downarrow}\sin\frac{\theta_{\rho}-\theta_{\sigma}}{\sqrt{2}}\right) \nonumber\\
& & -\frac{\Delta}{2\pi a}\gamma_{\uparrow}\gamma_{\downarrow}\Gamma_{\uparrow}\Gamma_{\downarrow}\cos\sqrt{2}\theta_{\sigma}
    +\frac{\tilde{\Delta}}{2\pi v}i\gamma_{\uparrow}\gamma_{\downarrow} \frac{i \partial_{\tau}\theta_{\sigma}}{\sqrt{2}} \Bigg].
\end{eqnarray}
Note the appearance of the new marginal term described by coupling constant $\tilde{\Delta}$.

We now perform a perturbative RG analysis up to the second-order in coupling coefficients. The details of the calculations are presented in Appendix \ref{app:RG2order_SU2_BreakingBoundary}. Here we summarize our results for $K_{\sigma}=1$:
\begin{eqnarray}\label{eq:RG2}
 \frac{dt}{dl} & = & \left(\frac{3}{4}-\frac{1}{4K_{\rho}}-\frac{\Delta}{4\pi v } \right)t -\frac{\tilde{\Delta} \tilde{t}}{2\pi v } , \label{eq:U1B_RG_t} \\
\frac{d\tilde{t}}{dl} & = & \left(\frac{3}{4}-\frac{1}{4K_{\rho}}+\frac{\Delta}{4\pi v }\right)\tilde{t} -\frac{ \tilde{\Delta} t }{2\pi v } , \label{eq:U1B_RG_tt}\\
\frac{d\Delta}{dl} & = & -\left(\frac{1}{K_{\rho}}-1\right) \frac{t^{2}-\tilde{t}^{2}}{4 \pi v},\label{eq:U1B_RG_Delta} \\
\frac{d\tilde{\Delta}}{dl} & = & - B(K_{\rho}) \frac{t \tilde{t}}{4 \pi v} .\label{eq:U1B_RG_tDelta}
\end{eqnarray}
The generation of the $\Delta$ term (proportional to $t^2-\tilde{t}^2$) originates from the processes involving two different spin channels of the lead whereas
the generation of the $\tilde{\Delta}$ term (proportional to $t\tilde{t}$ ) comes from processes within the same spin channel. Both of these terms can be generated {\it only} in the presence of the interaction in the lead. This fact follows from the definition of the
function $B(K_{\rho})$
\begin{equation}
 B(K_{\rho}) = \frac{ C(1/2K_\rho-1/2) }{C(1/2)\, C(1/2K_\rho)} \left(\frac{1}{K_{\rho}}+1 \right)>0.
\end{equation}
Here the function $C(\nu)$ is defined as
\begin{align}\label{eq:Cnu}
C(\nu)=\lim_{\delta \rightarrow 0^+} \int_0^{\infty} \frac{e^{-\delta z}\cos(z)}{(z+1)^\nu}dz,
\end{align}
 and originates from the integration over relative coordinate, $\tau-\tau'$ during the RG procedure, see Appendix \ref{app:RG2order_SU2_BreakingBoundary}.
In the non-interacting limit, $K_\rho\rightarrow 1$, $C(\nu\rightarrow 0^{+})\propto \nu$, and thus, the RG equation for $\tilde\Delta$ becomes
\be \label{eq:RGeqforDelta}
\frac{d\tilde{\Delta}}{dl}  \approx -\frac{c_5}{4\pi v}\left(\frac{1}{K_{\rho}}-1\right) t \widetilde{t},
\ee
where numerical constant $c_5\approx 11.5$.
As mentioned, both $\Delta$ and $\tilde{\Delta}$ cannot be generated in the RG in the absence of interactions in the lead (i.e. $K_\rho=1$).

Using Eqs.\eqref{eq:RG2} it is instructive to analyze first the flow in the non-interacting limit, in which case $\Delta=\tilde{\Delta}=0$. Both $t$ and $\tilde{t}$ are relevant and flow to strong coupling. As follows from the discussion in the previous section,
the exact boundary condition at the Andreev reflection fixed point is determined by the initial values of $t$ and $\tilde{t}$ and
we can identify the corresponding limits by looking at the scattering matrix, i.e. $t \gg \tilde t $ corresponds
to $\psi_{\sigma}(0)=-\psi^\dag_{\sigma}(0)$,   $t \ll \tilde t $ corresponds to $\psi_{\sigma}(0)=\psi^\dag_{\sigma}(0)$ and finally
$t = \tilde t $ corresponds to $\psi_{\sigma}(0)=-i\psi^\dag_{-\sigma}(0)$, see Fig. \ref{fig:RGflow}a.

We now analyze the RG flow for not-too-strong repulsive interactions $1/3\lesssim K_\rho<1$.
First of all, one can notice that even if we start with initial conditions $\Delta(l_0)=0$, $\tilde{\Delta}(l_0)=0$,
the corresponding four-fermion terms are going to be generated by the RG procedure. Here $l_0$ is initial length cutoff.
Since the couplings $\Delta$ and $\tilde{\Delta}$ affect the RG flow differently, we now have 4-parameter phase diagram.
Based on the perturbative RG equations, one can see that both $t$ and $\tilde{t}(l)$ will grow under RG, see Fig. ~\ref{fig:RGflow}. Thus, normal reflection fixed point is unstable in this parameter regime.

\subsubsection{RG analysis near spin-flip Andreev reflection fixed point $\mathbb{SFA}$}

We now analyze the stability of the spin-flip Andreev reflection fixed point $\mathbb{SFA}$ which corresponds to the following boundary conditions:
\begin{eqnarray}\label{eq:SFA_BC}
  \psi_{L\up}(0) &=& -i\psi^\dagger_{R\dn}(0), \\
  \psi_{L\dn}(0) &=& -i\psi^\dagger_{R\up}(0).
\end{eqnarray}
In the bosonization language, the bosonic fields $\phi_{\sigma}(0)=0$ and $\theta_{\rho}(0)=-\pi/(2\sqrt{2})$ are pinned,
and the Klein factors have the relation $\Gamma_{\uparrow L}=\Gamma_{\downarrow R}$ and
$\Gamma_{\downarrow L}=\Gamma_{\uparrow R}$. Now, let us study all the fermion bilinear perturbations at the boundary allowed by TR symmetry. First of all, one can show that the normal backscattering is vanishing in this case, in agreement with the scattering calculation in Sec.~\ref{sec:IIB1}.
Indeed, using the boundary conditions~\eqref{eq:SFA_BC} one can show that
\begin{eqnarray}
 &&\psi_{L\uparrow}^{\dagger}(0)\psi_{\uparrow,R}(0)+ \psi_{R\downarrow}^{\dagger}(0)\psi_{L\downarrow}(0) + h.c.\nonumber \\
 &&=-i\psi_{L\uparrow}^{\dagger}(0)\psi_{L\downarrow}^{\dagger}(0) + i\psi_{L\uparrow}(0)\psi_{L\downarrow}(0) +h.c. = 0
\end{eqnarray}
Note that for s-wave spin-singlet superconductor the boundary conditions are different:
$\psi_{L\up}(0)=\mp i\psi^\dagger_{R\dn}(0)$ and $\psi_{L\dn}(0)=\pm i\psi^\dagger_{R\up}(0)$, and the backscattering term
$\sim \sin \sqrt{2}\phi_{\rho}$ does not vanish. Since this term is relevant for $K_\rho<1$, the Andreev reflection fixed point is unstable in an s-wave superconductor-LL junction.

Let's now consider allowed Andreev reflection bilinear processes. Among those, the only allowed bilinear term is spin-conserving Andreev reflection:
\begin{eqnarray}
 H_{1B}^{SFA}&=&\lambda_1^{SFA}(\psi_{L\uparrow}^{\dagger}\psi_{\uparrow,R}^{\dagger} + \psi_{R\downarrow}^{\dagger}\psi_{L\downarrow}^{\dagger} + h.c.) \nonumber\\
 &=& \lambda_1^{SFA}(i\psi_{L\uparrow}^{\dagger}\psi_{L\downarrow} + i\psi_{R\downarrow}^{\dagger}\psi_{\uparrow,R} + h.c.) \nonumber\\
 &=& 2\frac{\lambda_1^{SFA}}{2\pi a} \left(i\Gamma_{L\up}\Gamma_{L\dn}+i\Gamma_{R\dn}\Gamma_{R\up}\right)\cos\sqrt{2}\theta_{\sigma}.
\end{eqnarray}
Additionally, we also consider the following four-fermion term
\begin{eqnarray}
 H_{2B}^{SFA}&=&\frac{\lambda_2^{SFA}}{(2\pi a)^2}(\psi_{L\up}^{\dagger}\psi_{R\up}\psi_{L\dn}^{\dagger}\psi_{R\dn}+h.c.)\nonumber\\
  &=& 2\lambda_2^{SFA} \;\Gamma_{L\up}\Gamma_{R\up}\Gamma_{L\up}\Gamma_{L\dn}\Gamma_{R\dn}\cos2\sqrt{2}\phi_{\rho},
\end{eqnarray}
which corresponds to two-electron backscattering. The leading order perturbative RG equations for $\lambda_1^{SFA}$
and $\lambda_2^{SFA}$ are give by
\begin{eqnarray}
 \frac{d\lambda_1^{SFA}}{dl}&=& \left(1-\frac{1}{K_{\sigma}}\right)\lambda_1^{SFA}, \\
 \frac{d\lambda_2^{SFA}}{dl}&=& (1-4K_{\rho})\lambda_2^{SFA}.
\end{eqnarray}
One can see that the first term $\lambda_1^{SFA}$ is marginal for $\mathbb{SU}(2)$ symmetric Luttinger liquid lead $K_{\sigma}=1$,
whereas the second coupling becomes relevant for $K_{\rho}<1/4$ indicating that the $\mathbb{SFA}$ fixed point becomes unstable
for strong repulsive interactions. If the $SU(2)$ spin symmetry is broken in the lead, the $\mathbb{SFA}$ fixed point
becomes unstable for $K_{\sigma}>1$, and the system will flow towards the $\mathbb{A\times A}$ fixed point.
On the other hand, the $\mathbb{SFA}$ is stable for $K_{\sigma}<1$.

\subsubsection{RG analysis near spin-conserving Andreev fixed point $\mathbb{A \times A}$}

As shown in Sec.~\ref{sec:IIB1}, the boundary conditions near $\mathbb{A\times A}$ fixed point are $\psi_{L,s}(0) = e^{i\alpha}\psi_{R,s}^{\dagger}(0)$ with $\alpha=0$ or $\pi$. Thus, the boson fields are $\theta_{\rho}=\pm\pi/\sqrt{2}$ and $\theta_{\sigma}=0$ are pinned at the boundary, and the Klein factors satisfy the relations
$\Gamma_{L,s}=\Gamma_{R,s}=\Gamma_{R}$. In the $U(1)$-conserving case, there are only
irrelevant perturbations for $1/3<K_{\rho}<1$ such as two-electron backscattering
\begin{align}
 H_{2B}^{A\times A}=&\lambda_2^{A\times A} \psi_{L\uparrow}^{\dagger}(0)\psi_{R\uparrow}(0)\psi_{L\downarrow}^{\dagger}(0)\psi_{R\downarrow}(0)+h.c. \nonumber \\
&= \frac{\lambda_2}{(2\pi a)^2}\sin(2\sqrt{2}\phi_{\rho}).
\end{align}
Additionally, if $U(1)$ symmetry is broken, the spin-flip Andreev
reflection processes are allowed
\begin{eqnarray}
 H_{1B}^{A\times A}&=& \lambda_1^{A\times A}(\psi_{R,\uparrow}^{\dagger}(0)\psi_{L,\downarrow}^{\dagger}(0)-\psi_{R,\downarrow}^{\dagger}(0)\psi_{L,\uparrow}^{\dagger}(0))+h.c.\nonumber\\
 &=& 4 i \Gamma_{\uparrow} \Gamma_{\downarrow} \sin \sqrt{2} \phi_{\sigma}.
\end{eqnarray}
The leading order perturbative RG equations for $\lambda_1^{A\times A}$
and $\lambda_2^{A\times A}$ are give by
\begin{eqnarray}
 \frac{d\lambda_1^{A\times A}}{dl}&=& (1-K_{\sigma})\lambda_1^{A\times A}, \\
 \frac{d\lambda_2^{A\times A}}{dl}&=& (1-4K_{\rho})\lambda_2^{A\times A}.
\end{eqnarray}
One can see that the first term $\lambda_1^{A\times A}$ is marginal for $SU(2)$ symmetric Luttinger liquid lead $K_{\sigma}=1$,
whereas the second coupling becomes relevant for $K_{\rho}<1/4$ indicating that $\mathbb{A\times A}$ fixed point becomes unstable
for strong repulsive interactions. If the $SU(2)$ spin symmetry is broken in the lead, the $\mathbb{A\times A}$ fixed point
becomes unstable for $K_{\sigma}<1$, and the system will flow towards the $\mathbb{SFA}$ fixed point.
On the other hand, the $\mathbb{A \times A}$ is stable for $K_{\sigma}>1$. Exactly at $K_\sigma=1$, both $\lambda_1^{A\times A}$ and $\lambda_1^{SFA}$ terms are marginal and compete with each other. Thus, generically both spin-conserving and spin-flip Andreev reflection processes will be present and their relative strength depends on microscopic details. This conclusion is consistent with the non-interacting results ($K_{\rho}=1$) discussed in Sec.\ref{sec:IIB1}. Our main results are summarized in Fig. \ref{fig:RGflow}.

\section{Majorana Kramers pair - Quantum dot - Normal lead junction}\label{sec:QD}
\subsection{Theoretical model}\label{sec:QDmodel}

In this section we study effect of local electron-electron interactions and consider the system consisting of a QD with a single spin-degenerate level coupled to MKP $\gamma_{\up,\dn}$, localized at the end of a TRITOPS, and a NL. The schematic plot of the device is shown in Fig. \ref{fig:device} b). Assuming that TR symmetry and $\mathbb{U}(1)$-spin rotation symmetry are preserved and the induced gap in the topological superconductor is sufficiently larger than other energy scales of the problem, the low-energy effective Hamiltonian of the system can be written as
\begin{align}
H&=\sum_\sigma\epsilon d^\dagger_\sigma d_\sigma + U n_\up n_\dn + V +H_{NL}\\
V&=\sum_\sigma [i\lambda_\sigma\gamma_\sigma (d_\sigma + d^\dagger_\sigma)+t_\sigma(d^\dagger_\sigma \psi_\sigma(0)+\text{h.c.})]
\end{align}
where $d^\dagger_\sigma$ and $d_\sigma$ are creation and annihilation operators on the QD, $n_\sigma=d^\dagger_\sigma d_\sigma$, $\epsilon$ is the chemical potential of the QD, $U$ is the strength of the electron-electron interaction on the QD, $\psi^\dagger_\sigma$ and $\psi_\sigma$ are fermion creation and annihilation operators in the NL, and $t(\lambda_\sigma)$ is the tunneling coefficient between the NL(MKP) and the QD. For the perturbative RG analysis, we adopted the same Hamiltonian for NL as Eq. \eqref{eq:Hlead} with $K=1$. For slave-boson mean-field theory analysis, we assumed quadratic dispersion $\xi_k$ for the NL. We set $t_\sigma$ and $\lambda_\sigma$ to be real. Time-reversal symmetry requires $t_\up=t_\dn=t$ and $\lambda_\up=-\lambda_\dn=\lambda$. The Hamiltonian $H_{NL}$ represents semi-infinite NL ($x\geq0$) with hopping $t_0$. We are interested in the limit where $\epsilon<0$, $U+\epsilon>0$ such that the QD favors single occupation, and weak coupling regime $|t|,|\lambda|\ll \min(-\epsilon,U-\epsilon)$. We also consider the non-interacting limit for NL. In this limit, one can simplify the effective Hamiltonian by projecting it onto single-occupation subspace~\cite{wolff}, see Appendix \ref{app:SW} for details. The effective Hamiltonian becomes $H=H_{\text{NL}}+H_b$ with the boundary Hamiltonian $H_b$ being
\begin{widetext}
\begin{align}
H_b&=\xi_+\bigg[\frac{t^2}{2}\vec{S}\cdot\vec{s}(0)+\frac{i\lambda^2}{2}\gamma_\up\gamma_\dn S_y  +\frac{i\lambda t}{2} \bigg(\gamma_\up(\psi_\up+\psi^\dagger_\up)S_z+ \gamma_\dn(\psi_\dn+\psi^\dagger_\dn)S_z +\gamma_\up(\psi_\dn S^- +\psi^\dagger_\dn S^+)-\gamma_\dn(\psi_\up S^+ +\psi^\dagger_\up S^-) \bigg) \bigg]\nonumber\\
&+\xi_-\left[\frac{i\lambda t}{2}\left(\gamma_\up(\psi_\up+\psi^\dagger_\up)-\gamma_\dn(\psi_\dn+\psi^\dagger_\dn)\right)\right],
\label{eq:Hb}
\end{align}
\end{widetext}
where $\vec{S}=d^\dagger_\alpha \boldsymbol{\sigma}_{\alpha\beta} d_\beta$, $\vec{s(0)}=\psi^\dagger_\alpha(0) \boldsymbol{\sigma}_{\alpha\beta} \psi_\beta(0)$, $S^+=S_x+iS_y$, $S^-=S_x-iS_y$ and the coefficients $\xi_\pm$ are defined as
\be
\xi_\pm=\frac{1}{|\epsilon|}\pm\frac{1}{U-|\epsilon|}.
\ee
In the limit $\lambda \rightarrow0$, the first term $\sim t^2$ drives the system to the Kondo fixed point where a spin in QD and a spin in the lead form a spin-singlet state. In the presence of the Majorana coupling $\lambda$, additional terms appear in the Hamiltonian. These Majorana-induced couplings favor the strong-correlation between QD spin and MKP, and, therefore, compete with Kondo coupling.

The critical difference between the present Hamiltonian~\eqref{eq:Hb} and that of time-reversal broken case with single Majorana mode, for example in Ref. \onlinecite{Kondo-Majorana}, is the presence of the second term $\sim\lambda^2$. This time-reversal preserving interaction term between QD and MKP replaces the Zeeman-like coupling in the single Majorana mode case. While the Zeeman-like coupling becomes zero at the particle-hole symmetric point in the previous study\cite{Kondo-Majorana}, this interaction term is proportional to $\xi_+$ and is always non-zero for any position of the level $\epsilon$ in the dot. {Therefore, one cannot apply the same method as in Ref.~[\onlinecite{Kondo-Majorana}] to find the exact solution at the particle-hole symmetric point.} To understand low-energy properties of the system, we present below the results from two complementary calculations:  perturbative RG analysis and slave-boson mean-field theory in the limit of an infinite on-site repulsion.

\subsection{Weak coupling perturbative RG analysis}\label{sec:QDRG}
In order to understand the effect of Majorana induced couplings on the infrared(IR) fixed point the system flows to, we study RG flow of the boundary couplings in the weak-coupling limit.
First, we introduce the following rescaled couplings: $M(l_0)=\xi_+\lambda^2$, $T_1(l_0)=\lambda t\xi_-$, $T_2(l_0)=\lambda t \xi_+$ and $J(l_0)=t^2\xi_+$. After performing standard bosonization procedure and rescaling the parameters, we obtain the following effective action at the boundary:
\begin{align}
S_{b}&=\int \frac{d\tau}{2\pi a} \bigg\{ iM\gamma_\uparrow\gamma_\downarrow S_y -\frac{iaJ^zS_z}{\sqrt{2} v}\partial_\tau\theta_\sigma \nonumber\\
+&iT_1 \left[ \gamma_\up\Gamma_\up\cos\left(\frac{\theta_\rho+\theta_\sigma}{\sqrt{2}}\right)- \gamma_\dn\Gamma_\dn\cos\left(\frac{\theta_\rho-\theta_\sigma}{\sqrt{2}}\right) \right] \nonumber\\
+&iT_2^z S_z \left[ \gamma_\up\Gamma_\up \cos\left(\frac{\theta_\rho+\theta_\sigma}{\sqrt{2}}\right)+\gamma_\dn\Gamma_\dn \cos\left(\frac{\theta_\rho-\theta_\sigma}{\sqrt{2}}\right) \right] \nonumber\\
+&iT_2^\bot \bigg[ \gamma_\up\Gamma_\dn\left(S_x\cos \left(\frac{\theta_\rho-\theta_\sigma}{\sqrt{2}}\right)+S_y\sin\left(\frac{\theta_\rho-\theta_\sigma}{\sqrt{2}}\right)\right) \nonumber\\
-&\gamma_\dn\Gamma_\up\left( S_x \cos\left(\frac{\theta_\rho+\theta_\sigma}{\sqrt{2}}\right)-S_y\sin\left(\frac{\theta_\rho+\theta_\sigma}{\sqrt{2}}\right)\right)\bigg]\nonumber\\
-&iJ^\bot \Gamma_\up\Gamma_\dn \left(S_x\sin{\sqrt{2}\theta_\sigma}+S_y\cos\sqrt{2}\theta_\sigma \right) \bigg\}
\label{eq:Sb0}
\end{align}
Here we have introduced couplings $T_2^{z,\bot}$ and $J^{z,\bot}$ for convenience. We will recover spin-rotation symmetry by setting $T_2^z=T_2^\bot$ and $J^z=J^\bot$ at the end of the calculation. We will focus on the non-interacting limit for NL, but adding small repulsive interaction in NL does not change our conclusion.

Let us now perform perturbative RG analysis up to the second order in couplings near ultraviolet normal reflection fixed point. The procedure of the calculations is similar to the one presented in Appendix \ref{app:RG2order_SU2} and \ref{app:RG2order_SU2_BreakingBoundary}. The RG equations for the couplings read
\ba
\frac{dM}{dl}&=&M+\frac{T_2^2}{\pi v}\\
\frac{dT_1}{dl}&=&\frac{T_1}{2}\\
\frac{dT_2}{dl}&=&\frac{T_2}{2}+\frac{T_2J}{\pi v}\\
\frac{dJ}{dl}&=&\frac{J^2}{\pi v}
\ea
From these RG equations, we can see that $M$ is the most relevant coupling while the Kondo coupling is only marginally relevant. Thus, the system generically flows to the Majorana strong coupling fixed point.
If initially $M(l_0) \ll J(l_0)$, the system can still reach the Kondo strong coupling fixed point.
One can estimate the crossover scale, $\lambda_c$, by solving $M(l^*)=J(l^*)\sim 1$ ($l^*$ is the crossover length scale) which leads to the following estimate for the critical coupling
\be
\lambda_c \sim \frac{1}{\xi_+} \exp \left(-\frac{\pi v}{2 \xi^+ t^2}\right),
\label{eq:lambdac}
\ee
which defines a crossover between the two regimes. In deriving this estimate, we have ignored the second order contributions from $T_2^2$ term assuming that it is small.

Around the Kondo strong coupling fixed point, the spin operator in QD is absorbed by NL and, therefore, acquires scaling dimension one. As a result, $T_2$  and $M$ terms become irrelevant and marginal, respectively. However, the term proportional to $T_1$ is still relevant and drives the system to Andreev reflection strong coupling fixed point with $\mathbb{A\times A}$ boundary condition and $G(0)=4e^2/h$, see Sec.~\ref{sec:SU2NoRashba}.

Let us now study the nature of Majorana strong coupling fixed point defined by  $M(l^*)\sim 1$ and  $J(l^*) \ll 1$. The two degenerate (Kramers) states that minimize $iM\gamma_\uparrow\gamma_\downarrow S_y$ term are
\ba
\ket{\psi_1}&=&\ket{i\gamma_\up\gamma_\dn=-1, S_y=1}\nonumber\\\,\\
\ket{\psi_2}&=&\ket{i\gamma_\up\gamma_\dn=1, S_y=-1}.\nonumber\\\nonumber
\ea
Assuming that $M$ is large, one can project the rest of the boundary terms on to this low-energy manifold and simplify the boundary problem. Since the ground state is an eigenstate of $S_y$ and $i\gamma_\up\gamma_\dn$, the terms that are proportional to $\gamma\otimes I$ and $\gamma\otimes S_y$ will be projected to zero. The remaining boundary terms at particle-hole symmetric point (i.e. $T_1=0$) are
\begin{align}\label{eq:Hmb}
\!\!H_{\rm mb}
&=i T_2(l^*)\left[ \beta_\up(\psi_\up+\psi_\up^\dagger)-\beta_\dn(\psi_\dn+\psi_\dn^\dagger)\right] \nonumber\\
&+\frac{iJ(l^*)}{2} \beta_\up\beta_\dn(-i\psi_\up^\dagger\psi_\dn+i\psi_\dn^\dagger\psi_\up),
\end{align}
where introduced generalized Majorana operators $\beta_\up=(\gamma_\up S_z -\gamma_\dn S_x)/2$ and $\beta_\dn=-(\gamma_\dn S_z +\gamma_\up S_x)/2$. In the ground state manifold with $i\gamma_\up\gamma_\dn S_y=-1$, these new operators behave as MKP, $\{\beta_\sigma,\beta_{\sigma'}\}=2\delta_{\sigma,\sigma'}$. Now one can notice that the effective Hamiltonian~\eqref{eq:Hmb} is exactly the same as that in Eq.~\eqref{eq:model0} with $K_\rho=1$ and $\Delta_{\text{AN}}=0$. Therefore, using the results from the previous section and the condition $T_2(l^*)\gg J(l^*)$, we can immediately conclude that the system will flow to the strong coupling fixed point under RG and will be governed by the $\mathbb{A\times A}$ boundary condition $\psi_\sigma(0)=-\psi_\sigma^\dagger(0)$.

As follows from the aforementioned analysis, the coupling of QD to MKP leads to a non-trivial many-body ground-state where the spin on the QD gets entangled with the fermion parity of the MKP. Due to the change in the boundary conditions for lead electrons, the zero-bias tunneling conductance is $G=4 e^2/h$ due to perfect Andreev reflection phenomenon. Further insight about the physical properties of the system can be obtained using a complementary approach - slave-boson mean field theory.

\subsection{Slave-boson mean field theory}\label{sec:QDSB}
In this section, we develop a slave-boson mean-field approach for an infinite repulsive interaction in QD (i.e. $U \rightarrow \infty$). In this case one can completely exclude double occupancy state from the Hilbert space, see, for example, Ref. \cite{Nagaosa_book}. Next, one can represent the creation and annihilation operators for the QD as $d^\dagger_\sigma\rightarrow f^\dagger_\sigma b$ and $d_\sigma \rightarrow f_\sigma b^\dagger$ with an additional constraint $b^\dagger b+\sum_\sigma f^\dagger_\sigma f_\sigma=1$ where $b$ is a boson operator representing an empty state. Thus, the effective action of the system in terms of new fields variables reads
\begin{widetext}
\begin{align}
S_{\text{sb}}=&\int d\tau \sum_\sigma \bigg[\sum_k\psi^*_{k,\sigma}(\partial_\tau+\xi_k)\psi_{k,\sigma}+f^*_\sigma(\partial_\tau+\epsilon)f_\sigma + i\lambda_\sigma\gamma^1_\sigma(f_\sigma b^* + f^*_\sigma b)+\sum_k t(f^*_\sigma\psi_{k,\sigma} b+\psi^*_{k,\sigma}f_\sigma b^*)\nonumber\\
&+\frac{1}{2}\sum_{i=1,2}\gamma^i_\sigma\partial_\tau \gamma^i_\sigma +i\delta_{1\sigma}\gamma^1_\sigma\gamma^2_\sigma+i\delta_2 \gamma^1_\sigma\gamma^2_{-\sigma} +\eta\left(\frac{b^* b-1}{2}+f^*_\sigma f_\sigma\right)\bigg]
\label{eq:mfe0},
\end{align}
\end{widetext}
where $\eta$ is the Lagrange multiplier, $\gamma^1$ and $\gamma^2$ correspond to the Majorana modes at the end of the TSC near the QD and at the opposite end. $\delta_{1\up}=-\delta_{1\dn}=\delta_1$ and $\delta_2$ mix the Majorana modes at the opposite ends for finite size system.

\subsubsection{Mean-field solution}
We now develop self-consistent mean-field theory for the problem. We first calculate mean-field solution for the action~\eqref{eq:mfe0} and replace boson fields with their mean-field value $\langle b \rangle= \langle b^* \rangle=b$ and solve for $b$ and $\eta$. Here, without loss of generality, we assumed that $b$ is real since the phase can be gauged away.  In the next section, we will study effect fluctuations around the mean-field saddle point and specify precisely the meaning of the mean-field solution for this low-dimensional model.

\begin{figure}[t]
	\centering
  \includegraphics[scale=0.45]{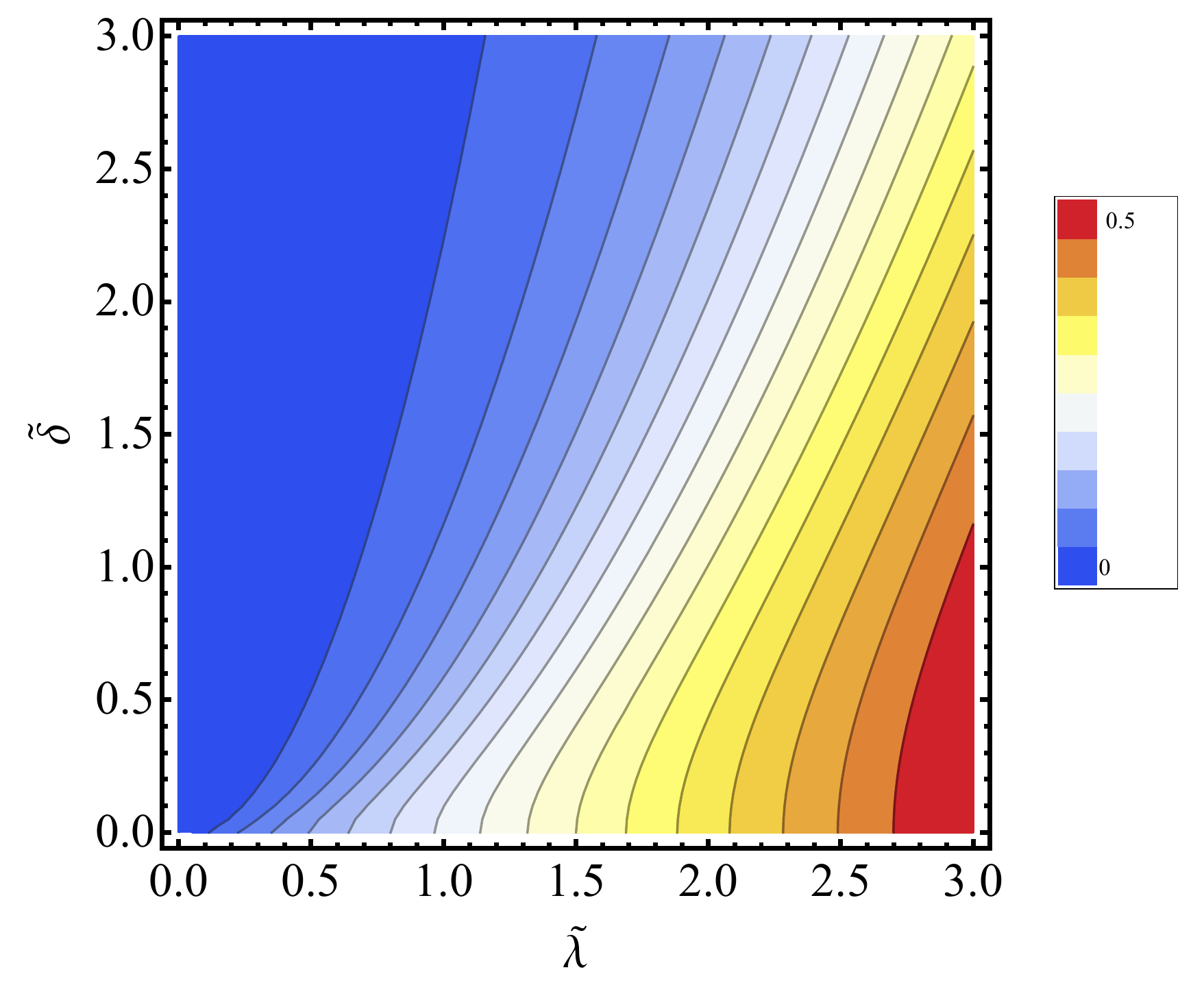}
\caption{The solution $b$ of the mean-field equation as a function of $\tilde\lambda$ and $\tilde\delta$. We set $\epsilon=-6\Gamma$ and $\Lambda=50\Gamma$.}
\label{fig:mfs2}
\end{figure}

The mean-field equations can be obtained by minimizing the action~\eqref{eq:mfe0}:
\ba
\frac{\partial S}{\partial \eta}&=&b^2+\sum_{\sigma}\langle f^*_\sigma f_\sigma \rangle -1=0 \label{eq:mfe1}\\
\frac{\partial S}{\partial b}&=&2b\eta+t\sum_{k,\sigma}(\langle f^*_\sigma\psi_{k,\sigma}\rangle + \langle \psi^*_{k,\sigma} f_\sigma \rangle)\nonumber\\
&&+i \sum_\sigma \lambda_\sigma \langle \gamma^1_\sigma(f^*_\sigma + f_\sigma)\rangle =0 \label{eq:mfe2}
\ea
The details of the calculation of the correlation functions are presented in the Appendix \ref{app:SB1}. We first consider the limit $T, \delta_1,\delta_2 \rightarrow 0$ and assume that $|\epsilon|\gg |\lambda|,\Gamma$ where $\Gamma=\pi \nu_F|t|^2$ such that the probability for empty state in QD $b^2$ is small. In this limit, the first equation becomes
\be
\epsilon+\eta\approx \frac{\pi}{2}\Gamma b^4,
\ee
Substituting $\eta\approx -\epsilon$ back into Eq.\eqref{eq:mfe2} and neglecting smaller terms, one finds
\be
\eta-\frac{2\Gamma}{\pi}\ln\frac{\Lambda}{\Gamma b^2}-\frac{|\lambda|}{\sqrt{2}b}\approx 0\label{eq:sbr}
\ee
where we assumed $\lambda\gg\Gamma b$.
For $\lambda\rightarrow 0$ we recover the solution for the Kondo-dominated regime:
\ba
T_K&\equiv&\Gamma b^2=\Lambda e^{-\frac{\pi|\epsilon|}{2\Gamma}}.
\ea
If Majorana coupling $\lambda \gg \lambda_c$, $b$ is determined by the last term in Eq. \eqref{eq:sbr}:
\be
b\approx\frac{|\lambda|}{\sqrt{2}|\epsilon|}.
\ee
The crossover between two regimes occurs at
\be
\lambda_c \approx \sqrt{\frac{2\Lambda}{\Gamma}}|\epsilon|e^{-\frac{\pi|\epsilon|}{4\Gamma}}
\ee
which qualitatively agrees with the estimate for $\lambda_c$ from the RG analysis, see Eq.~\eqref{eq:lambdac}.
In the presence of the Majorana splitting $\delta_1$ and $\delta_2$ and for arbitrary value of $\lambda/\Gamma b$, we can solve the mean-field equations numerically. In terms of  $\delta_1^2+\delta_2^2\equiv\delta^2$,  the second mean-field equation\eqref{eq:mfe2} now becomes
\be
\frac{|\epsilon|}{\Gamma}-\frac{2}{\pi}\ln\frac{\Lambda}{\Gamma b^2}-2I(b,\tilde\lambda,\tilde\delta)=0,\label{eq:sbr2}
\ee
where
\begin{align}
I(b,&\tilde\lambda,\tilde\delta)=\frac{b^2\tilde\lambda^2}{\pi}\times\\
&\int_0^\infty dx \frac{x(x-\tilde\delta^2-b^2\tilde\lambda^2)}{(x+b^4)(x(x-\tilde\delta^2-2b^2\tilde\lambda^2)^2+b^4(x-\tilde\delta^2)^2)} \nonumber
\end{align}
One can numerically solve the Eq. \eqref{eq:sbr2} for self-consistent solution $b$ as a function of $\tilde\lambda=\lambda/\Gamma$ and $\tilde\delta=\delta/\Gamma$, see Fig. \ref{fig:mfs2} for results. One can see that if we increase the splitting for Majoranas $\delta$, the Kondo correlations become more important and eventually start to dominate. As a result, the magnitude of $\lambda_c$ defining the crossover between two different fixed points is increased.

\subsubsection{Gaussian fluctuations around mean-field solution}\label{sec:gaussianfluctuation}
In the previous section, we found mean-field solution for $b$ and $\eta$ of Eq.~\eqref{eq:mfe0}. We now analyze the stability of the mean-field solution with respect to fluctuations. This issue is rather subtle, and has been discussed extensively in the context of the Kondo problem~\cite{Coleman'87}. Indeed, one can check that the action~\eqref{eq:mfe0} is invariant with respect to local gauge transformations $b \rightarrow b e^{i\theta}$ and $f\rightarrow e^{i\theta}f$. The mean-field solution appears to break this $\mathbb{U}(1)$ symmetry. However, as we will show below, the fluctuations will restore this symmetry.

We now make a transformation to the ``radial coordinates" and rewrite $b(\tau)=s(\tau)e^{i\theta(\tau)}$. One can check that the action~\eqref{eq:mfe0} is invariant with respect to local gauge transformations $b \rightarrow b e^{i\theta}$ and $f\rightarrow e^{i\theta}f$ and $\eta(\tau)\rightarrow \eta+i\partial_\tau\theta$. Therefore, we can expand the action in terms of fluctuations $\delta s(\tau)$ and $\partial_\tau \theta(\tau)$
\be
s(\tau)=\bar{s}+\delta s(\tau) , \,\,\,\,\,\, \eta(\tau)=\bar\eta+i\partial_\tau\theta(\tau).
\ee
around the corresponding saddle point. Here $\bar{s}$ is the mean-field solution for $b$, defined in the previous section.
After integrating out fermions, the effective action reads
\be
S_{\text{eff}}=-\Tr\ln\left[ \mathcal{G}^{-1}(s,\eta) \right]+\int d\tau \left[\eta(s^2-1)+s\partial_\tau s \right]\label{eq:effS}
\ee
where
\be
\mathcal{G}^{-1}(s,\eta)=\left(
\begin{array}{cc}
-G_{f}^{-1} & sG_{\gamma} s\\
sG_{\gamma}s & -\tilde{G}_{f}^{-1}
\end{array}
\right)
\ee
and
\begin{align}
G_{f}&=-\frac{1}{\partial_\tau+\epsilon+\eta+s(G_\psi+G_{\gamma})s}, \\
\tilde{G}_{f}&=-\frac{1}{\partial_\tau-\epsilon-\eta+s(\tilde{G}_\psi+G_{\gamma})s}, \\
G_\psi&=-\sum_k\frac{t^2}{\partial_\tau+\xi_k},  \,\,\,\,\, \tilde{G}_\psi=-\sum_k\frac{t^2}{\partial_\tau-\xi_k},  \\
G_{\gamma}&=-\frac{\lambda^2}{\partial_\tau+\delta}.
\end{align}
We now expand the fields $s$ and $\eta$ around their mean-field values and collect the quadratic terms in $\delta s$ and $\partial_\tau \theta$. In Matsubara frequency domain, the effective action for the Gaussian fluctuation can be written as
\be
S^{(2)}_{\text{eff}}=\frac{1}{2\beta}\sum_\nu (\dot\theta_{-\nu} \,\, \delta s_{-\nu}) \left(\begin{array}{cc}
 \Gamma^{\dot\theta \dot\theta}_\nu &  \Gamma^{\dot\theta s}_\nu \\
\Gamma^{\dot\theta s}_\nu & \Gamma^{ss}_\nu
\end{array}\right) \left(\begin{array}{c}
\dot\theta_\nu \\
\delta s_\nu
\end{array}\right)
\ee
where we introduced Fourier transform
\be
\delta s(\tau) =\frac{1}{\beta}\sum_\nu \delta s_\nu e^{-i\omega_\nu \tau}, \,\,\,\,\, \dot\theta(\tau) =\frac{1}{\beta}\sum_\nu \dot\theta_\nu e^{-i\omega_\nu \tau}
\ee
with bosonic Matsubara frequency $\omega_\nu=2\pi\nu/\beta$. The details of the calculation of the above matrix elements $\Gamma^{ij}$ are given in Appendix \ref{app:SB2}. Here we simply highlight the main results. We first note that $\Gamma^{\dot\theta s}_\nu\approx 2i\bar{s} $ near the mean-field solution $\bar{\eta} \approx -\epsilon$. Diagonal element $\Gamma^{ss}_{\nu}$ and $\Gamma^{\theta \theta}_{\nu}$ can be obtained using the analytic continuation of fermionic Matsubara frequency $i\omega_n\rightarrow \omega$ and integrating around the two branch cuts $\text{Im}[\omega]=0$ and $\text{Im}[\omega]=-\omega_\nu$. The correlation function of $\delta s$ and $\dot\theta$ is given by
\ba
D_{\dot\theta \dot\theta}(i\omega_{\nu})&=&\frac{\Gamma^{ss}_\nu}{\Gamma^{\dot\theta \dot\theta}_\nu\Gamma^{ss}_\nu+4\bar{s}^2},\\
D_{ss}(i\omega_\nu)&=&\frac{\Gamma^{\dot\theta \dot\theta}_\nu}{\Gamma^{\dot\theta \dot\theta}_\nu\Gamma^{ss}_\nu+4\bar{s}^2},
\ea
and govern the dynamics of the fluctuating fields $\delta s(\tau)$ and $ \dot\theta(\tau)$. We can now address the question regarding the restoration of the broken $\mathbb{U}(1)$ symmetry.

\begin{figure}[t]
\includegraphics[width=3.4in]{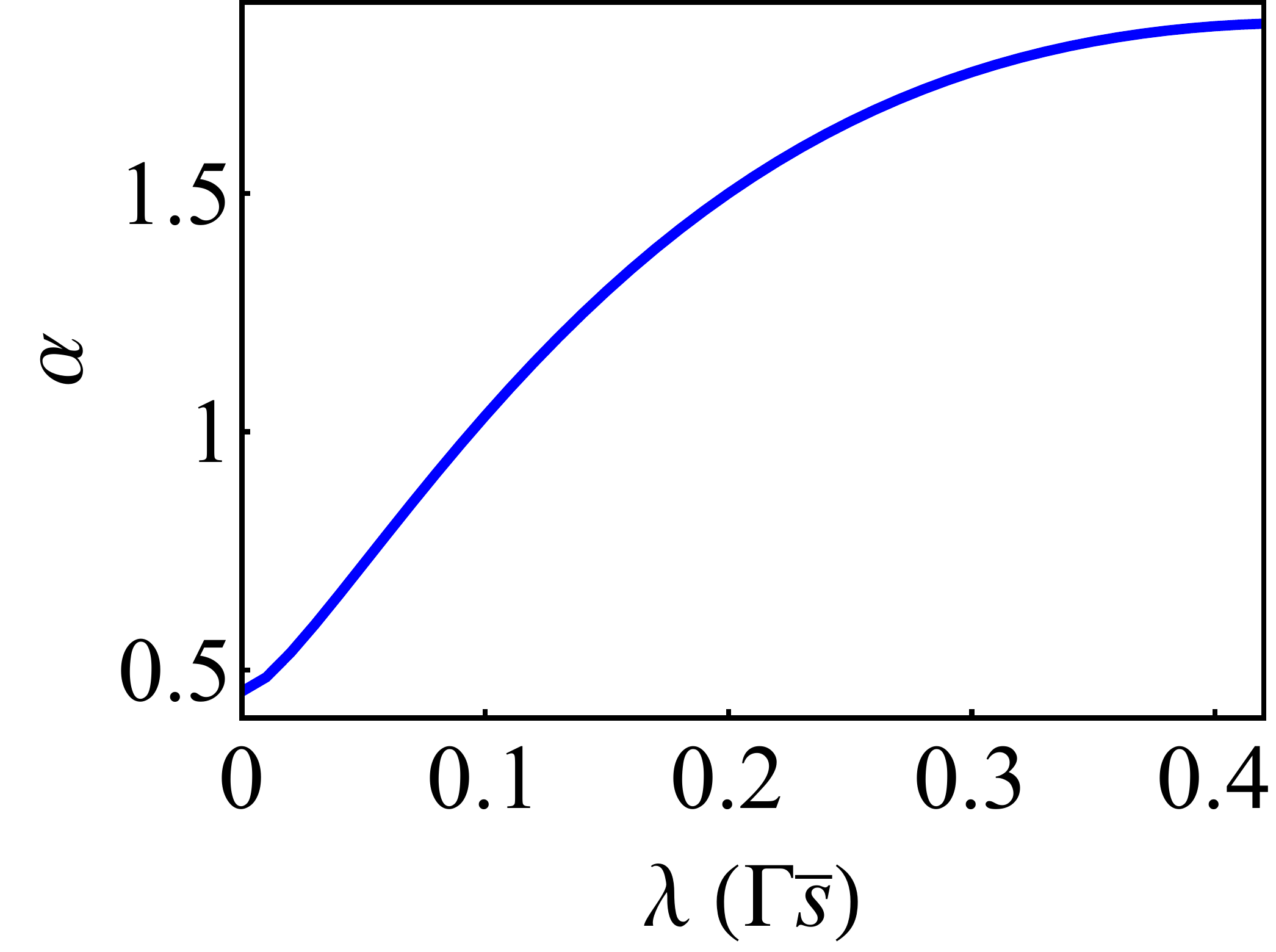}
\caption{The exponent $\alpha$ as a function of Majorana coupling strength $\lambda$.}
\label{fig:exponent}
\end{figure}

Let us consider the correlation function $\langle b(\tau)b^*(0)\rangle$. The mean-field solution assumes that  $\langle b(\tau)b^*(0)\rangle \rightarrow \bar{s}^2$ for $\tau \rightarrow \infty$. It has been shown, however, in Ref.~\cite{Coleman'87} that above correlation function for the generalized Anderson model decays as a power-law $\langle b(\tau)b^*(0)\rangle\propto |\tau|^{-\alpha}$ with some non-universal exponent.
\begin{figure}[t]
\includegraphics[width=3.5in]{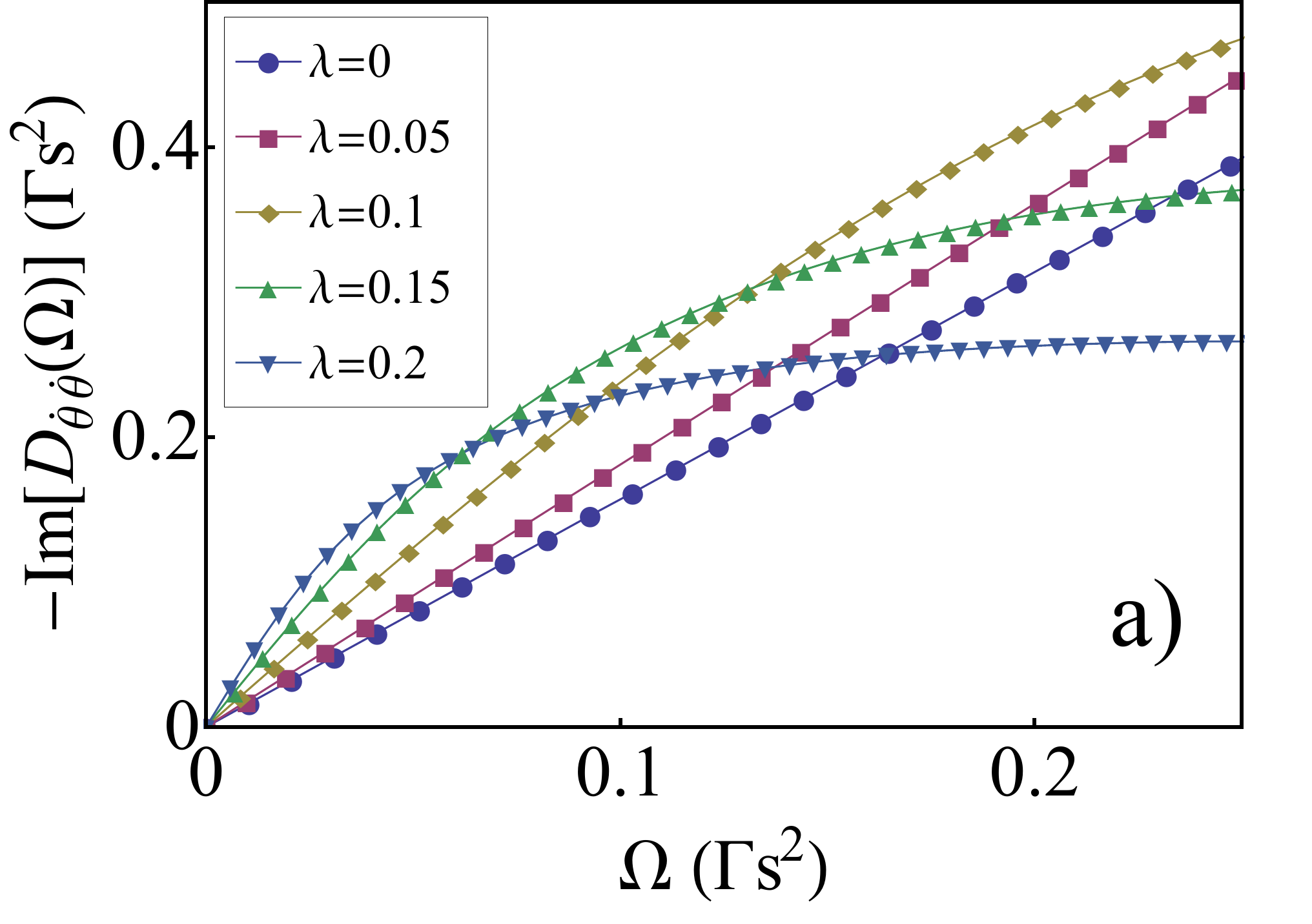}\\
\includegraphics[width=3.5in]{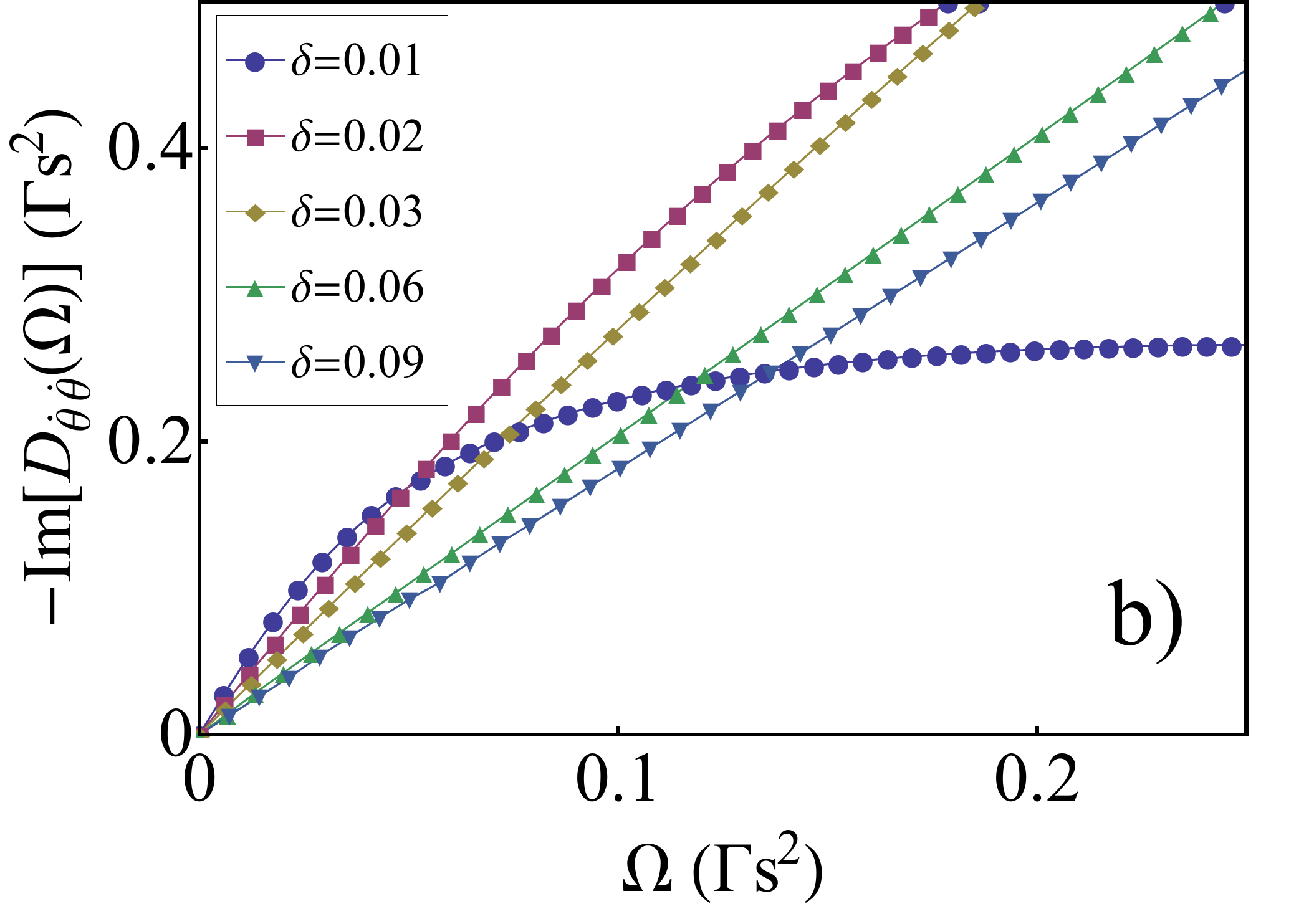}
\caption{The function $\text{Im}[D_{\dot\theta\dot\theta}(\Omega)]$ for different values of $\lambda$ and $\delta$. Here panels a) and b) correspond to  $\delta=0.01$ and $\lambda=0.2$; we used $\Gamma=1$, $\epsilon=-5$, $\Lambda=50$ here. }
\label{fig:ImGss}
\end{figure}
We now perform a similar analysis for QD-MKP problem at hand. Since $\langle s(\tau)s(0)\rangle\sim \bar{s}^2$ in the long time limit, one can decouple amplitude and phase fluctuations
\ba \label{eq:theta_corr}
\langle b(\tau)b^*(0)\rangle &\approx& \bar{s}^2  \langle e^{i(\theta(\tau)-\theta(0))}\rangle\nonumber\\
&=&\bar{s}^2 \exp(-\frac{1}{2}\langle[\theta(\tau)-\theta(0)]^2\rangle).
\ea
We can evaluate the exponent, following Ref.\cite{Coleman'87}, as
\begin{align}
\frac{1}{2}\langle[\theta(\tau)&-\theta(0)]^2\rangle=\frac{1}{\beta}\sum_{\nu\neq0} \frac{D_{\dot\theta\dot\theta}}{\omega_\nu^2}(1-e^{-i\omega_\nu\tau})\label{eq:bcontour}\\
&=-\oint\frac{d\Omega}{2\pi i}\frac{1-e^{-\Omega\tau}}{1-e^{-\beta \Omega}}\frac{D_{\dot\theta\dot\theta}(\Omega)}{\Omega^2}\\\nonumber
&\stackrel{T\rightarrow 0 }{=}-\int^\infty_0 \frac{d\Omega}{\pi}\frac{1-e^{-\Omega\tau}}{\Omega^2}\text{Im}[\lim_{\xi\rightarrow 0^+}D_{\dot\theta\dot\theta}(\Omega+i\xi)].\nonumber
\end{align}
Here Matsubara sum was evaluated by integrating along the branch cut $\text{Im}[\Omega]=0$ using the analytic continuation for bosonic Matsubara frequency $i\omega_\nu\rightarrow \Omega$. We find that $\text{Im}[D_{\dot\theta \dot\theta}(\Omega+i\epsilon)] \propto -\alpha\Omega$ in low frequency limit, see Fig. \ref{fig:ImGss}.
Here we eventually take $\epsilon \rightarrow 0$. Thus, the correlation function
\begin{equation}
 \langle b(\tau)b^*(0)\rangle \propto \tau^{-\alpha}
\end{equation}
decays as a power law, which is a key result of this section. In this sense, the situation is analogous to slave-boson theory for the Kondo problem.
The expression for $\alpha$ as a function of $\lambda_0=\lambda/\Gamma\bar{s}$ in the limit of zero splitting for MKP, $\delta\rightarrow 0$, is given by
\be
\alpha=\frac{1}{2}\frac{1}{(h(\lambda_0)+\bar{s}^2\pi^2/4)}
\ee
\be
h(\lambda)\!=\!\begin{cases}
-\frac{\ln[4\lambda^4]}{8\lambda^2}\!-\!\frac{1-4\lambda^2}{8\lambda^2\sqrt{1\!-\!8\lambda^2}}\ln\left[\frac{1\!-\!4\lambda^2+\sqrt{1-8\lambda^2}}{1-4\lambda^2-\sqrt{1-8\lambda^2}} \right], &\lambda<\frac{1}{2}\\
-\frac{\ln[4\lambda^4]}{8\lambda^2}\!-\!\frac{1-4\lambda^2}{4\lambda^2\sqrt{1\!-\!8\lambda^2}}\left(\frac{\pi}{2}\!-\!\tan^{-1}\frac{1-4\lambda^2}{\sqrt{1-8\lambda^2}}\right),  & \lambda \geq \frac{1}{2}
\end{cases} \nonumber
\ee
Using the corresponding mean-field solution of Eq. \eqref{eq:mfe2}, one can evaluate the exponent $\alpha$, see Fig. \ref{fig:exponent}. We find that the exponent $\alpha$ moderately increases with $\lambda$. When the Majorana splitting energy $\delta$ becomes larger, $\alpha$ decreases and eventually approaches the value in the Kondo limit $\alpha =\frac{1}{2}+O(\bar{s}^2)$.

Overall, we find that the correlation function~\eqref{eq:theta_corr} decays as a power law in the long-time limit which is qualitatively similar to phase fluctuations in the Kondo problem~\cite{Coleman'87}. This is the main result of this section showing that fluctuations ultimately restore $U(1)$ symmetry, in agreement with the Mermin-Wagner theorem, but the correlation function decays slowly in comparison with the ``disordered" high-temperature limit. The situation is reminiscent of quasi-long range order where the fluctuations ultimately restore broken symmetry but, at the same time, there is a well-defined mean-field amplitude of fluctuations (i.e. $\bar{s}\neq 0$) which opens up a gap in the spectrum.

\subsection{Differential tunneling conductance}

Using the mean-field theory developed in the previous sections, one can now calculate transport properties of the NL-QD-TSC junction. To compute the differential conductance $G$, one needs to compute scattering matrix of the system within the mean-field approximation. The slave-boson mean-field Hamiltonian can be written as
\ba
H_{sb}&=&H_{NL}+\sum_\sigma\bigg[\sum_{k}tb(f^\dagger_\sigma \psi_{k,\sigma}+\psi^\dagger_{k,\sigma}f_\sigma)+\tilde\epsilon f^\dagger_\sigma f_\sigma\nonumber\\
&& +i\lambda_\sigma b\gamma^1_\sigma(f^\dagger_\sigma+f_\sigma)+i\delta_{1\sigma}\gamma^1_\sigma\gamma^2_\sigma+i\delta_2\gamma^1_\sigma\gamma^2_{-\sigma} \bigg].
\ea
The scattering matrix for electrons close to the Fermi level is given by
\be
S(E)=1+2\pi i\hat{W}^\dagger(H_{local}-E-\pi i \hat{W}\hat{W}^\dagger)^{-1}\hat{W},
\ee
where $H_{local}$ is the Hamiltonian describing the ``local impurity" and $\hat{W}\propto tb$ is the matrix of coupling constants between local degrees of freedom and lead electrons.

Using the scattering matrix one can compute the probability for Andreev reflection and ultimately obtain differential conductance $G(V)$. In agreement with the analysis in Sec. \ref{sec:QDRG}, we find that zero-bias differential conductance is quantized $G(0)=4e^2/h$. In the limit of small bias voltage and zero splitting $\delta\rightarrow0$, the differential conductance  $G(V)$ reads
\be
G(V)\approx\frac{4e^2}{h}\,\frac{\Gamma_{\rm eff}^2}{\Gamma_{\rm eff}^2+(eV)^2}
\ee
with the width of the zero-bias peak changing from $\Gamma_{\text{eff}}\approx\text{min}\{ T_K, \frac{2\lambda^2}{\Gamma} \}$ in Kondo-dominated to $\Gamma_{\text{eff}}\approx\frac{\Gamma\lambda^2}{2\epsilon^2}$ in the Majorana-dominted regime.

In addition to the differential conductance, the signatures of MKP should be observable in shot noise and full counting statistics measurements as have been discussed in the context of a quantum dot coupled to a single Majorana zero mode, see, e.g., Refs.~\onlinecite{QDShotNoise,QDFCS}.

\section{Conclusions}\label{sec:conclusion}

 We study two new boundary impurity problems involving MKPs: Luttinger liquid - MKP and NL - QD - MKP junctions, see Fig. \ref{fig:device}. The presence of MKPs in these systems leads to a drastic change of their physical properties. A well-known example is the change of the differential tunneling conductance through such a junction from essentially zero (perfect normal reflection) to $4e^2/h$ (perfect Andreev reflection)~\cite{wong&Law12, Li15}.  This result, however, was obtained for non-interacting systems, and we extend the analysis to interacting systems.

For the first example (i.e. Luttinger liquid - MKP junction) we consider electron-electron interactions in the bulk of the wire. We find that perfect Andreev reflection fixed point is stable with respect to weak repulsive interactions in the lead. This result should be contrasted with the conventional  Luttinger liquid - s-wave superconductor junction where weak repulsive interactions destabilize Andreev reflection fixed point and drive the system back to the normal reflection fixed point~\cite{Fidkowski'12}. The reason for such a difference is the relative sign change in Andreev boundary conditions indicating that Luttinger liquid - MKP junction is similar, in this sense, to Luttinger liquid coupled to spin-triplet p-wave superconductor rather than spin-singlet s-wave superconductor. We perform perturbative RG analysis near perfect normal reflection and perfect Andreev reflection fixed points and propose a phase diagram, see Fig.~\ref{fig:RGflow}.

Next we investigate effect of local repulsive interactions in the normal lead-quantum dot- TRI topological superconductor junction. We show that the system flows to a new fixed point which is characterized by a strong entanglement of a QD spin with a MKP. These correlations ultimately lead to the change of boundary conditions for lead electrons: from Kondo to perfect Andreev boundary conditions. Using a combination of a perturbative RG analysis and slave-boson mean-field theory we identify the ground-state of the system and calculate tunneling conductance through the junction, demonstrating that zero-temperature differential tunneling conductance is $4e^2/h$.
As we increase Majorana coupling $\lambda$, the width of the zero-bias peak exhibits a crossover from the Kondo temperature $T_K$ to $\Gamma \lambda^2/\epsilon^2$ in the Majorana-dominated regime.
We have also studied effect of quantum fluctuations near the slave-boson mean-field saddle point and demonstrated that the mean-field solution is well-defined (in the quasi-long range order sense) and thus can be used to calculate the spectrum in the QD as well as other observables.

Our work represents the first step in understanding the signatures of Majorana Kramers pairs (class DIII superconductor) when coupled to an interacting lead. As we have shown, the phase diagram in this case is much richer than in the case of a single Majorana mode (class D superconductor), see Ref.~\cite{Fidkowski'12}. Our perturbative approach does not capture the crossover regime between perfect normal reflection and perfect Andreev reflection. It would be interesting to understand the phase diagram for $1/4 <K_{\rho}\lesssim 1/3$ using numerical methods and check our conjectured phase diagram.

In this paper we also analyzed Gaussian fluctuations around the slave-boson mean-field solution, see Sec. \ref{sec:gaussianfluctuation}.
We confirmed the stability of the low-energy mean-field solution by calculating the correlation function of the slave bosons. It would be interesting to study other physical quantities numerically such as impurity spectral
function and magnetic susceptibility of the impurity.

{\it Note added}. While this manuscript was in preparation, we became aware of related independent work on this subject~\cite{Pikulin2016} which has some overlap with Sec.~\ref{sec:MKPLL}.

\section{Acknowledgements}

We are grateful to Dmitry Pikulin, Jian Li and Andrei Bernevig for stimulating discussions. YK was supported by the Samsung Scholarship. RL wishes to acknowledge the hospitality of the Aspen Center for Physics and the support under NSF Grant \#1066293.
EG, JP and KF acknowledges the support from the Danish Council for Independent Research $|$ Natural Sciences.


\newpage

\appendix
\begin{widetext}

\section{Second order perturbative RG calculation for the case with $U(1)$ symmetry}\label{app:RG2order_SU2}

In this Appendix we provide details for the perturbative RG calculation for the case with $U(1)$ symmetry. We will use momentum shell RG procedure and calculate each term that is generated in the second order of perturbation theory.

In order to obtain the quadratic corrections to the RG flow Eq. (\ref{eq:U1RG_Delta}) and Eq. (\ref{eq:U1RG_delta}) of the main text,
let us consider the contribution from the $t_{\uparrow}t_{\downarrow}$ term:
\begin{align}
\delta S^{(tt)} & = -\frac{1}{2}\int d\tau\int d\tau'\frac{t_{\uparrow}t_{\downarrow}}{(2\pi a)^2}\gamma_{\uparrow}(\tau)\Gamma_{\uparrow}(\tau)
\;\gamma_{\downarrow}(\tau')\Gamma_{\downarrow}(\tau')\\
 & \quad\quad\quad\times\left(\left \langle\cos\frac{\theta_{\rho}(\tau)+\theta_{\sigma}(\tau)}{\sqrt{2}}\cos\frac{\theta_{\rho}(\tau')-\theta_{\sigma}(\tau')}{\sqrt{2}}\right \rangle_>-\left \langle\cos\frac{\theta_{\rho}(\tau)+\theta_{\sigma}(\tau)}{\sqrt{2}}\right \rangle_>\left \langle\cos\frac{\theta_{\rho}(\tau)-\theta_{\sigma}(\tau)}{\sqrt{2}}\right \rangle_>\right). \nonumber
\end{align}
Here $\langle \ldots \rangle_>$ denotes integrating out the fast modes, $\Lambda/b < |\omega| < \Lambda$, where $b=e^l \approx 1+dl$ describes the change in UV cutoff under RG procedure. One can evaluate above correlation functions using the following identity $\langle e^{\frac{i}{\sqrt{2}}\theta_{j}^{>}(\tau)}\rangle=e^{-\frac{1}{4}\langle\theta_{j}^{>}(\tau)^{2}\rangle}$. Taking into account that the correlation function  $\langle(\theta^<_j(\tau)-\theta_j^<(\tau'))^2\rangle_<$ decays sufficiently quickly with $\tau-\tau'$, one can use the following approximation:
\be
e^{i(\theta^<_j(\tau)-\theta^<_j(\tau'))}\approx(1+(\tau-\tau')\partial_\tau\theta^<_j+\ldots)e^{-\frac{1}{2}\langle(\theta^<_j(\tau)-\theta_j^<(\tau'))^2\rangle_<}
\ee
where the correlation functions are given by
\ba
\langle(\theta_j(\tau)-\theta_j(\tau'))^2\rangle&=&\langle(\theta^<_j(\tau)-\theta_j^<(\tau'))^2\rangle_<+\langle(\theta^>_j(\tau)-\theta_j^>(\tau'))^2\rangle_>=\frac{2}{K_j}\ln \left[ \frac{a}{v|\tau-\tau'|+a}\right] \\
g_j(\tau-\tau')&\equiv&\langle\theta^>_j(\tau)\theta^>_j(\tau')\rangle_>=\frac{1}{K_j}\int^\Lambda_{\Lambda/b} \frac{d\omega}{\omega} \cos(\omega|\tau-\tau'|).
\ea

After some manipulations, one finds that
\begin{align}
&\left \langle\cos\frac{\theta_{\rho}(\tau)+\theta_{\sigma}(\tau)}{\sqrt{2}} \cos\frac{\theta_{\rho}(\tau')-\theta_{\sigma}(\tau')}{\sqrt{2}} \right \rangle_> \nonumber\\
=&\frac{1}{2}\left(\frac{\cos\frac{\theta_{\sigma}^{<}(\tau)+\theta_\sigma^<(\tau')}{\sqrt{2}}}{(\Lambda|\tau-\tau'|+1)^{\frac{1}{2K_\rho}}}e^{-\frac{1}{2}(g_\sigma(0)+g_\sigma(\tau-\tau'))}+\frac{\cos\frac{\theta_{\rho}^{<}(\tau)+\theta_\rho^<(\tau')}{\sqrt{2}}}{(\Lambda|\tau-\tau'|+1)^{\frac{1}{2K_\sigma}}} e^{-\frac{1}{2}(g_\rho(0)+g_\rho(\tau-\tau'))}\right).
\end{align}
The contribution of disconnected part is given by
\begin{align}
\left \langle\cos\frac{\theta_{\rho}(\tau)+\theta_{\sigma}(\tau)}{\sqrt{2}}\right \rangle_>& \left \langle \cos\frac{\theta_{\rho}(\tau)-\theta_{\sigma}(\tau)}{\sqrt{2}}\right \rangle_>\\
=&\frac{1}{2}\left(\frac{\cos\frac{\theta_{\sigma}^{<}(\tau)+\theta_\sigma^<(\tau')}{\sqrt{2}}}{(\Lambda|\tau-\tau'|+1)^{\frac{1}{2K_\rho}}}e^{-\frac{1}{2}(g_\sigma(0)+g_\rho(\tau-\tau'))}+\frac{\cos\frac{\theta_{\rho}^{<}(\tau)+\theta_\rho^<(\tau')}{\sqrt{2}}}{(\Lambda|\tau-\tau'|+1)^{\frac{1}{2K_\sigma}}} e^{-\frac{1}{2}(g_\rho(0)+g_\sigma(\tau-\tau'))}\right).
\end{align}
Before we proceed, it is important to note that
\begin{equation}
 g_{j}(\tau-\tau')\equiv\langle\theta_{j}^{>}(\tau)\theta_{j}^{>}(\tau')\rangle =
 \frac{1}{K_j}\int_{\Lambda/b}^{\Lambda} \frac{d\omega}{\omega} \cos[\omega (\tau-\tau')]\approx \frac{1}{K_j}\cos[\Lambda(\tau-\tau')] dl
\end{equation}
and thus $g_{j}(0)\approx dl/K_{j}$. Let's introduce new variables: center-of-mass $T=\frac{\tau+\tau'}{2}$ and relative coordinates $s=\tau-\tau'$.
The correction to the action to the linear order of $dl$ becomes
\ba
\delta S^{(tt)}  \!&= & \!\frac{1}{4}\frac{t_{\uparrow}t_{\downarrow}}{(2\pi a)^2}\int_{0}^\infty dT \int_{-\infty}^\infty ds \gamma_{\uparrow}\gamma_{\downarrow}
     \Gamma_{\uparrow}\Gamma_{\downarrow}\cos\frac{\theta_{\sigma}(T+s/2)+\theta_{\sigma}(T-s/2)}{\sqrt{2}} \left( \frac{1}{2K_\rho}-\frac{1}{2K_\sigma} \right) \left( \frac{\cos(\Lambda s)}{(\Lambda |s|+1)^{\frac{1}{2K_\rho}}}\right)dl \nonumber\\
&\!+&\!\frac{1}{4}\frac{t_{\uparrow}t_{\downarrow}}{(2\pi a)^2} \int_{0}^\infty\! dT \int_{-\infty}^\infty ds \gamma_{\uparrow}\gamma_{\downarrow}\Gamma_{\uparrow}
   \Gamma_{\downarrow} \cos\frac{\theta_{\rho}(T+s/2)+\theta_{\rho}(T-s/2)}{\sqrt{2}} \left( \frac{1}{2K_\sigma}-\frac{1}{2K_\rho} \right) \left( \frac{\cos(\Lambda s)}{(\Lambda|s|+1)^{\frac{1}{2K_{\sigma}}}} \right)dl
\ea
Since the above expression has a power law decay in $\Lambda |s|$, the contributions to the integral comes from the short time $|s|\sim 1/\Lambda$. After the simplification, the total contribution to the effective action reads
\ba
 \delta S^{(tt)} & \approx & \frac{1}{4}\frac{t_{\uparrow}t_{\downarrow}}{(2\pi a)^2}\frac{2 dl}{\Lambda}  \left(\frac{1}{2K_{\rho}}-\frac{1}{2K_{\sigma}}\right) C\left(\frac{1}{2K_{\rho}}\right)  \int_{0}^{\beta}dT\gamma_{\uparrow}\gamma_{\downarrow}\Gamma_{\uparrow}
 \Gamma_{\downarrow}\cos\sqrt{2}\theta_{\sigma}(T)\nonumber\\
 && +\frac{1}{4}\frac{t_{\uparrow}t_{\downarrow}}{(2\pi a)^2}\frac{2 dl}{\Lambda} \left(\frac{1}{2K_{\sigma}}-\frac{1}{2K_{\rho}}\right) C\left(\frac{1}{2K_{\sigma}}\right) \int_{0}^{\beta}dT\gamma_{\uparrow}\gamma_{\downarrow}\Gamma_{\uparrow}\Gamma_{\downarrow}\cos\sqrt{2}\theta_{\rho}(T)
\label{eq:stt}
\ea
where the function $C(\nu)$ is defined as
\begin{equation}
 C(\nu) = \lim_{\delta \rightarrow 0^+} \int_{0}^{\infty} \frac{e^{-\delta x}\cos x }{(1+x)^{\nu}} dx.
 \label{eq:functionC}
\end{equation}
Notice that $C(\nu)$ is proportional to $\nu$ when $\nu \rightarrow 0$. Away from $v=0$ is simply $\mathcal{O}(1)$ constant which can be absorbed into the definition of the coupling constants.

Combining all the terms in Eq. \eqref{eq:stt}, we find the following contributions to the RG equations at quadratic order in t:
\ba
\frac{d\Delta^{(2)}}{dl}&=&-\frac{t^2}{4\pi v}\left(\frac{1}{K_\rho}-\frac{1}{K_\sigma}\right)\\
\frac{d\Delta_{\rm AN}^{(2)}}{dl}&=&\frac{t^2}{4\pi v}\left(\frac{1}{K_\rho}-\frac{1}{K_\sigma}\right),
\ea
see Eqs. (\ref{eq:U1RG_Delta}) and Eq. (\ref{eq:U1RG_delta}) of the main text.
Note that factor of 2 here originates  from the switching time coordinates $\tau$ and $\tau'$.

We now consider the contribution to RG equations from the crossed terms proportional to $t \Delta$, see Eq. (\ref{eq:U1RG_t}) in the main text. The relevant terms in the second order expansion of $S_T$ are
\begin{eqnarray}
\delta S^{(t\Delta)} & = & -\frac{1}{2}\int d\tau\int\tau'\frac{-i\, t_{\uparrow}\Delta}{\left(2\pi a\right)^{2}}\gamma_{\uparrow}(\tau)
\Gamma_{\uparrow}(\tau)\,\gamma_{\uparrow}(\tau')\gamma_{\downarrow}(\tau')\Gamma_{\uparrow}(\tau')\Gamma_{\downarrow}(\tau')\nonumber\\
 &  & \qquad\qquad\times\left(\langle\cos\frac{\theta_{\rho}(\tau)+\theta_{\sigma}(\tau)}{\sqrt{2}}\cos\sqrt{2}\theta_{\sigma}(\tau')\rangle_>-\langle\cos\frac{\theta_{\rho}(\tau)+\theta_{\sigma}(\tau)}{\sqrt{2}}\rangle_>\langle\cos\sqrt{2}\theta_{\sigma}(\tau')\rangle_>\right).
\end{eqnarray}
Given that $\langle\gamma_{s}(\tau)\gamma_{s}(\tau')\rangle={\rm sgn}(\tau-\tau')$
and $\langle\Gamma_{s}(\tau)\Gamma_{s}(\tau')\rangle={\rm sgn}(\tau-\tau')$, above correlation function can be simplified
\begin{equation}
\gamma_{\uparrow}(\tau)\Gamma_{\uparrow}(\tau)\,\gamma_{\uparrow}(\tau')\gamma_{\downarrow}(\tau')\Gamma_{\uparrow}(\tau')\Gamma_{\downarrow}(\tau')=\gamma_{\downarrow}(\tau')\Gamma_{\downarrow}(\tau').
\end{equation}
Next, we evaluate the bosonic part of the correlation function
\begin{align}
 \langle\cos\frac{\theta_{\rho}(\tau)+\theta_{\sigma}(\tau)}{\sqrt{2}}&\cos\sqrt{2}\theta_{\sigma}(\tau')\rangle_>-
 \langle\cos\frac{\theta_{\rho}(\tau)+\theta_{\sigma}(\tau)}{\sqrt{2}}\rangle_>\langle\cos\sqrt{2}\theta_{\sigma}(\tau')\rangle_>\nonumber\\
 & \approx \frac{1}{2K_\sigma} \cos\frac{\theta_{\rho}^{<}(T)-\theta_{\sigma}^{<}(T)}{\sqrt{2}}\frac{\cos(\Lambda s) dl}{(\Lambda|s|+1)^{\frac{1}{2K_{\sigma}}}}.
\end{align}
Here we dropped irrelevant terms generated by the RG procedure such as  $\cos\frac{\theta_{\rho}^{<}(\tau)+3\theta_{\sigma}^{<}(\tau)}{\sqrt{2}}$.
Using similar steps as in the previous section, we obtain the correction to the action proportional $t\Delta$:
\begin{equation}
\delta S^{(t\Delta)}\approx \frac{1}{4}\frac{i\, t_{\uparrow}\Delta}{\left(2\pi a\right)^{2}}\frac{2 dl}{\Lambda} \frac{1}{2K_{\sigma}}
C\left(\frac{1}{2K_{\sigma}}\right) \int_{0}^{\beta}dT\gamma_{\downarrow}(T)\Gamma_{\downarrow}(T)\cos\frac{\theta_{\rho}(T)-\theta_{\sigma}(T)}{\sqrt{2}}.
\end{equation}

Similarly, we evaluate the contribution to the effective action from $t_{\uparrow}\Delta_{\rm AN}$ term to find
\begin{eqnarray}
\delta S^{(t\Delta_{\rm AN})} & = & -\frac{1}{2}\int d\tau\int d\tau'\frac{-i\, t_{\uparrow}\Delta_{\rm AN}}{\left(2\pi a\right)^{2}}\gamma_{\uparrow}(\tau)
\Gamma_{\uparrow}(\tau)\,\gamma_{\uparrow}(\tau')\gamma_{\downarrow}(\tau')\Gamma_{\uparrow}(\tau')\Gamma_{\downarrow}(\tau')\nonumber\\
 &  & \qquad\qquad\times\left(\langle\cos\frac{\theta_{\rho}(\tau)+\theta_{\sigma}(\tau)}{\sqrt{2}}\cos\sqrt{2}\theta_{\rho}(\tau')\rangle
 -\langle\cos\frac{\theta_{\rho}(\tau)+\theta_{\sigma}(\tau)}{\sqrt{2}}\rangle\langle\cos\sqrt{2}\theta_{\rho}(\tau')\rangle\right)\nonumber\\
 & \approx & \frac{1}{4}\frac{i\, t_{\uparrow}\Delta_{\rm AN}}{\left(2\pi a\right)^{2}}\frac{2 dl}{\Lambda} \frac{1}{2K_{\rho}}
C\left(\frac{1}{2K_{\rho}}\right) \int_{0}^{\beta}dT\gamma_{\downarrow}(T)\Gamma_{\downarrow}(T)\cos\frac{\theta_{\rho}(T)-\theta_{\sigma}(T)}{\sqrt{2}}.
\end{eqnarray}
Once again here we dropped the irrelevant term $\cos\frac{3\theta_{\rho}^{<}(\tau)+3\theta(\tau)}{\sqrt{2}}$.
Combining all the terms in Eq. \eqref{eq:stt}, we find the quadratic part of the RG flow Eq. (\ref{eq:U1RG_t}) in the main text:
\begin{equation}
 \frac{dt}{dl}  = -\frac{\Delta t }{4\pi v K_{\sigma}} -\frac{\Delta_{\rm AN} t}{4\pi v K_{\rho}}.
\end{equation}

\section{Boundary conditions in the presence of spin-orbit coupling}\label{app:soc_boundary}

In order to perform the bosonization procedure, we first analyze the non-interacting Hamiltonian and calculate proper boundary conditions for fermion fields at the boundary $x=0$. The corresponding Schrodinger equation is given by ($\hbar=1$)
\begin{equation}
 \left[\! \left( -\frac{1}{2m}\frac{d^2}{dx^2}\!-\! \mu \!+\!V(x) \right) \hat{I} \!+\! \left( \alpha_R i \frac{d}{dx} \right)\hat{\sigma}_y  \right]
    \psi(x) = E \psi(x),
\end{equation}
where $\hat{I}$ is $2\times 2$ identity matrix and $\hat{\sigma}_i$ is Pauli matrix, the
potential $V(x)$ is given by
\begin{equation}
 V(x) = \begin{cases}
         V & \qquad \text{for} \; x<0 \\
         0 & \qquad \text{for} \; x>0 \\
        \end{cases},
\end{equation}
with $V \gg E_F$ and $E_F$ being the Fermi energy. The eigenfunctions can be written as
\begin{align}
 \psi_{I}(x>0) & = \phi_{\uparrow L} e^{ik_{1} x} |\uparrow\rangle + \phi_{\downarrow L} e^{ik_{2} x} |\downarrow\rangle \nonumber\\
       &  + \phi_{\uparrow R} e^{-ik_{2} x} |\uparrow\rangle + \phi_{\downarrow R} e^{-ik_{1} x} |\downarrow\rangle, \\
 \psi_{II}(x<0) &= A_{\uparrow} e^{\kappa_{\uparrow} x} |\uparrow\rangle +A_{\downarrow} e^{\kappa_{\downarrow} x} |\downarrow\rangle
\end{align}
where $|\uparrow/\downarrow\rangle$ are the eigenstates of $\hat{\sigma}_y$ (i.e. $\hat{\sigma}_y|\uparrow/\downarrow\rangle=\pm|\uparrow/\downarrow\rangle$)
and $\frac{\hbar^2 k_{1/2}^2}{2m}-\mu \mp \alpha_R k_{1/2} =E $; $\kappa_{\uparrow/\downarrow} = i \pm \frac{m\alpha_R}{\hbar}+\kappa_0$ with $\kappa_0 = \sqrt{V-\mu -E-\frac{m^2\alpha_R^2}{\hbar^2}}>0$.
We impose the following boundary conditions
\begin{eqnarray}
  \psi_{I}(0^+) &=& \psi_{II}(0^-) \nonumber\\ \, \\
  \left( \partial_x \mathbb{I} + i \frac{m\alpha_R}{\hbar} \hat{\sigma}_y \right) \psi_{I}(0^+)
       &=& \left( \partial_x \mathbb{I} + i \frac{m\alpha_R}{\hbar} \hat{\sigma}_y \right) \psi_{II}(0^-)\nonumber
\end{eqnarray}
(Note that the current operator in the presence of SOC is
$J = \frac{\partial H}{\partial P_x}=-i\frac{\hbar}{m} (\partial_x \mathbb{I} + i \frac{m\alpha_R}{\hbar} \hat{\sigma}_y)$),
and obtain
\begin{eqnarray}
 &&\frac{\phi_{\uparrow L}}{\phi_{\uparrow R}} = \frac{i(k_2 + \frac{m\alpha_R}{\hbar})+\kappa_0}{i(k_1 - \frac{m\alpha_R}{\hbar})-\kappa_0}= e^{i \zeta},\nonumber\\
 &&\frac{\phi_{\downarrow L}}{\phi_{\downarrow R}} = \frac{i(k_1 - \frac{m\alpha_R}{\hbar})+\kappa_0}{i(k_2 + \frac{m\alpha_R}{\hbar})-\kappa_0}= e^{i\zeta}.
\end{eqnarray}
where we use the condition $k_1 -  \frac{m\alpha_R}{\hbar} = k_2 + \frac{m\alpha_R}{\hbar}$.
Therefore,
the boundary condition is simply $\psi_{L\uparrow}(0)=e^{i\zeta} \psi_{\uparrow,R}(0)$ and $\psi_{L\downarrow}(0)= e^{i\zeta} \psi_{R\downarrow}(0)$
($R/L$ denotes right/left moving field, and $\uparrow/\downarrow$ is the spin index).
As long as the scattering at the boundary preserves helicity (i.e. boundary operator commutes with the Hamiltonian), the boundary conditions at $x=0$ correspond to perfect normal reflection of $\sigma_y$ eigenstates. Therefore, we can proceed by using standard Abelian bosonization procedure in the basis of $\sigma_y$-eigenstates.

\section{Second order perturbative RG calculation with broken $U(1)$ symmetry}\label{app:RG2order_SU2_BreakingBoundary}

In this section we evaluate additional terms contributing to the RG equations when $U(1)$ symmetry is broken.

We first consider the contribution of $\tilde{\Delta}t_{\downarrow}$ to the  RG flow Eq. (\ref{eq:U1B_RG_tt})in the main text:
\begin{eqnarray}
\delta S^{(\tilde{\Delta}t_{\downarrow})} & = & \frac{1}{2}\int d\tau\int d\tau'\frac{\tilde{\Delta}}{2\pi v}(-1)\gamma_{\uparrow}(\tau)\gamma_{\downarrow}(\tau)\frac{it_{\downarrow}}{2\pi a}\gamma_{\downarrow}(\tau')\Gamma_{\downarrow}(\tau')\nonumber\\
 &  & \times\frac{1}{2\sqrt{2}}\Big[\langle\partial_{\tau}\theta_{\sigma}(\tau)e^{\frac{i}{\sqrt{2}}\left(\theta_{\sigma}(\tau')-\theta_{\rho}(\tau')\right)}\rangle-\langle\partial_{\tau}\theta_{\sigma}(\tau)\rangle\langle e^{\frac{i}{\sqrt{2}}\left(\theta_{\sigma}(\tau')-\theta_{\rho}(\tau')\right)}\rangle\nonumber\\
 &  & \qquad+\langle\partial_{\tau}\theta_{\sigma}(\tau)e^{-\frac{i}{\sqrt{2}}\left(\theta_{\sigma}(\tau')-\theta_{\rho}(\tau')\right)}\rangle-\langle\partial_{\tau}\theta_{\sigma}(\tau)\rangle\langle e^{-\frac{i}{\sqrt{2}}\left(\theta_{\sigma}(\tau')-\theta_{\rho}(\tau')\right)}\rangle\Big]\nonumber\\
 & \approx & -\frac{b^{-\frac{1}{4}}}{8 \pi v}\frac{i\tilde{\Delta}t_{\downarrow}}{2\pi a}\, \int d T \gamma_{\uparrow}\Gamma_{\downarrow}\sin\frac{\theta_{\rho}^{<}-\theta_{\sigma}^{<}}{\sqrt{2}}\;\int ds\,\sgn(s)\partial_{s}g_{\sigma}(s)\nonumber\\
 & \approx & \frac{\tilde{\Delta}t_{\downarrow}}{2\pi a}\, \frac{ dl}{4\pi v K_\sigma} \int dT\; i\gamma_{\uparrow}\Gamma_{\downarrow}\sin\frac{\theta_{\rho}^{<}-\theta_{\sigma}^{<}}{\sqrt{2}}.
\end{eqnarray}
Here we use the definition $g_{j}(\tau-\tau')=\langle\theta_{j}^{>}(\tau)\theta_{j}^{>}(\tau')\rangle$with
$g_{j}(0)=\ln b/K_{j}$, and the following relations
\begin{eqnarray}
\langle\partial_{\tau}\theta_{\sigma}(\tau)e^{\pm\frac{i}{\sqrt{2}}\theta_{\sigma}(\tau')}\rangle & = & \partial_{\tau}\theta_{\sigma}^{<}(\tau)\; e^{\pm\frac{i}{\sqrt{2}}\theta_{\sigma}^{<}(\tau')}e^{-\frac{1}{4}\langle(\theta_{\sigma}^{>})^{2}\rangle}+e^{\pm\frac{i}{\sqrt{2}}\theta_{\sigma}(\tau')}\langle\partial_{\tau}\theta_{\sigma}^{>}(\tau)e^{\pm\frac{i}{\sqrt{2}}\theta_{\sigma}^{>}(\tau')}\rangle,\\
\langle\partial_{\tau}\theta_{\sigma}(\tau)\rangle\langle e^{\pm\frac{i}{\sqrt{2}}\theta_{\sigma}(\tau')}\rangle & = & \partial_{\tau}\theta_{\sigma}^{<}(\tau)\; e^{\pm\frac{i}{\sqrt{2}}\theta_{\sigma}^{<}(\tau')}e^{-\frac{1}{4}\langle(\theta_{\sigma}^{>})^{2}\rangle,}\\
\langle\partial_{\tau}\theta_{\sigma}^{>}(\tau)e^{\pm\frac{i}{\sqrt{2}}\theta_{\sigma}^{>}(\tau')}\rangle & = & \frac{\pm i}{\sqrt{2}}\partial_{\tau}\langle\theta_{\sigma}^{>}(\tau)\theta_{\sigma}^{>}(\tau')\rangle e^{-\frac{1}{4}\langle(\theta_{\sigma}^{>})^{2}\rangle}\\
\lim_{\delta\rightarrow 0^+}\int ds\,\sgn(s)\partial_{s}g_{\sigma}(s)e^{-\delta |s|} & = & -\frac{2}{K_\sigma}\ln b\approx -\frac{2 dl}{K_\sigma}.
\end{eqnarray}
At the end of the day, the correction to the RG equation reads
\begin{equation}
\frac{d\widetilde{t}_{\uparrow\downarrow}}{dl}=-\frac{\tilde{\Delta}t_{\downarrow}}{2\pi v K_{\sigma}}.
\end{equation}
Similarly, the contribution of $\tilde{\Delta}t_{\uparrow}$ will generate the following contribution:
the RG equation
\begin{equation}
\frac{d\widetilde{t}_{\downarrow\uparrow}}{dl}=-\frac{\tilde{\Delta}t_{\uparrow}}{2\pi v K_{\sigma}} .
\end{equation}
The cross term $\tilde\Delta \tilde{t}$ leads to the similar correction to $t$.
It is also straightforward to compute the contributions from $\Delta\tilde{t_i}$ terms using the same technique as in Appendix \ref{app:RG2order_SU2}.

We now evaluate the contribution of the $t\tilde{t}$ term in the second order expansion of $S_T$, see Eq. (\ref{eq:U1B_RG_tDelta}) in the main text. During this calculation we will encounter the expressions such as
$e^{i\frac{\theta_{\rho}(\tau)+\theta_{\sigma}(\tau)}{\sqrt{2}}}e^{-i\frac{\theta_{\rho}(\tau')+\theta_{\sigma}(\tau')}{\sqrt{2}}}$.
This term will contribute to the RG flow of $\tilde\Delta$. In order to demostrate this, one needs to carefully expand above expression up to the linear order in $s$:

\begin{eqnarray}
e^{i\frac{\theta_{\rho}(\tau)+\theta_{\sigma}(\tau)}{\sqrt{2}}}e^{-i\frac{\theta_{\rho}(\tau')+\theta_{\sigma}(\tau')}{\sqrt{2}}}&=&e^{i\frac{\theta_{\rho}(\tau)-\theta_{\rho}(\tau')}{\sqrt{2}}}e^{i\frac{\theta_\sigma(\tau)-\theta_{\sigma}(\tau')}{\sqrt{2}}}\\
&=&\left(1+s\frac{i\partial_T\theta_\rho}{\sqrt{2}}\right)\frac{1}{(\Lambda|s|+1)^{\frac{1}{2K_\rho}}}\left(1+s\frac{i\partial_T\theta_\sigma}{\sqrt{2}}\right)\frac{1}{(\Lambda|s|+1)^{\frac{1}{2K_\sigma}}}\\
&\sim&\left(\frac{1}{\Lambda|s|+1}+\frac{\sgn(s)}{\Lambda}\frac{i\partial_T(\theta_\rho+\theta_\sigma)}{\sqrt{2}}\right)\frac{1}{(\Lambda|s|+1)^{\frac{1}{2K_\rho}+\frac{1}{2K_\sigma}-1}}
\end{eqnarray}
After some algebra, one finds
\begin{eqnarray}
 \delta S^{t \widetilde{t}} &=& -\frac{1}{2}\int d\tau \int d\tau'\Bigg\{ \frac{t \widetilde{t}}{(2\pi a)^2} i\gamma_{\uparrow}(\tau)\Gamma_{\uparrow}(\tau) i\gamma_{\downarrow}(\tau')\Gamma_{\uparrow}(\tau')\nonumber\\
 &&  \qquad\qquad\qquad \times \Big[ \langle \cos\frac{\theta_{\rho}+\theta_{\sigma}(\tau)}{\sqrt{2}} \sin\frac{\theta_{\rho}+\theta_{\sigma}(\tau')}{\sqrt{2}} \rangle -\langle \cos\frac{\theta_{\rho}+\theta_{\sigma}(\tau)}{\sqrt{2}}\rangle\langle \sin\frac{\theta_{\rho}+\theta_{\sigma}(\tau')}{\sqrt{2}} \rangle\Big]\nonumber\\
 && \qquad\qquad\qquad+ \frac{t \widetilde{t}}{(2\pi a)^2} i\gamma_{\downarrow}(\tau)\Gamma_{\downarrow}(\tau) i\gamma_{\uparrow}(\tau')\Gamma_{\downarrow}(\tau')\nonumber\\
 &&  \qquad\qquad\qquad \times \Big[ \langle \cos\frac{\theta_{\rho}-\theta_{\sigma}(\tau)}{\sqrt{2}} \sin\frac{\theta_{\rho}-\theta_{\sigma}(\tau')}{\sqrt{2}} \rangle
-\langle \cos\frac{\theta_{\rho}-\theta_{\sigma}(\tau)}{\sqrt{2}}\rangle\langle \sin\frac{\theta_{\rho}-\theta_{\sigma}(\tau')}{\sqrt{2}} \rangle\Big]\Bigg\} \nonumber\\
 &\approx & \frac{1}{2}\int dT \int ds \frac{t \widetilde{t}}{(2\pi a)^2\Lambda} \gamma_{\uparrow}\gamma_{\downarrow} \frac{\partial_{\tau}\theta_{\sigma}}{\sqrt{2}}  \frac{\cos(s\Lambda)}{(\Lambda |s|+1)^{\frac{1}{2K_\rho}+\frac{1}{2K_\sigma}-1}} \left(\frac{1}{2K_\rho}+\frac{1}{2K_\sigma}\right)dl \nonumber\\
                           &\approx& -\frac{1}{8\pi v}\frac{t \widetilde{t}}{2\pi v} \left(\frac{1}{K_{\rho}}+\frac{1}{K_{\sigma}}\right) dl \,C\left(\frac{1}{2K_{\rho}}+\frac{1}{2K_\sigma}-1\right)\,\int dT i\gamma_{\uparrow}\Gamma_{\uparrow} \frac{i\partial_T\theta_{\sigma}}{\sqrt{2}}.
\end{eqnarray}
Once again we have to multiply the above expression by 2 due to the symmetry of between $\tau$ and $\tau'$.

\subsection{Perturbative RG equations near normal reflection fixed point: From  Eq. (\ref{eq:U1B_RG_t}) to Eq. (\ref{eq:U1B_RG_tDelta})}
Taking into account above results, one finds the following system of RG equations for generic values of $K_\rho$ and $K_\sigma$:
\begin{eqnarray}
 \frac{dt}{dl} & = & \left(1-\frac{1}{4K_{\rho}}-\frac{1}{4K_\sigma}-\frac{C(1/2K_\sigma)\Delta}{4\pi v K_\sigma} \right)t -\frac{\tilde{\Delta} \tilde{t}}{2\pi v K_\sigma} ,  \\
\frac{d\tilde{t}}{dl} & = & \left(1-\frac{1}{4K_{\rho}}-\frac{1}{4K_\sigma}+\frac{C(1/2K_\sigma) \Delta}{4\pi v K_\sigma }\right)\tilde{t} -\frac{ \tilde{\Delta} t }{2\pi v K_\sigma} , \\
\frac{d\Delta}{dl} & = & -\frac{C(1/2K_\rho) }{4 \pi v}\left(\frac{1}{K_{\rho}}-1\right) (t^{2}-\tilde{t}^{2}), \\
\frac{d\tilde{\Delta}}{dl} & = & -\frac{ C(1/2K_\rho+1/2K_\sigma-1) }{4 \pi v}\left(\frac{1}{K_{\rho}}+\frac{1}{K_\sigma}\right) t \tilde{t},
\end{eqnarray}
Compare with Eqs.~(\ref{eq:U1B_RG_t}) and (\ref{eq:U1B_RG_tDelta}) in the main text.
Provided the coefficients $C(x_i)$ are non-zero (i.e. $K_\rho, K_\sigma \neq 1/2$), one can rescale $C(1/2K_\sigma)\Delta\rightarrow \Delta$, $\sqrt{C(1/2K_\rho)C(1/2K_\sigma)}\, t \rightarrow t$ and $\sqrt{C(1/2K_\rho)C(1/2K_\sigma)}\, \tilde{t} \rightarrow \tilde t$ to absorb the $C(\nu)$'s in first three equations. Then the last equation becomes
\be
\frac{d\tilde{\Delta}}{dl}  =  -\frac{ C(1/2K_\rho+1/2K_\sigma-1) }{C(1/2K_\rho)C(1/2K_\sigma)} \left(\frac{1}{K_{\rho}}+\frac{1}{K_\sigma}\right)  \frac{t \tilde{t}}{4 \pi v},
\ee
and we recover Eq.~\eqref{eq:RGeqforDelta}.

\section{Derivation of the boundary Hamiltonian for Majorana-QD-LL system}\label{app:SW}

In this section we present details for the derivation of effective boundary Hamiltonian of our model. To derive the low-energy Hamiltonian we perform a Schrieffer-Wolff transformation\cite{wolff} and project out zero- and double-occupancy sectors. The projection operators to the n-occupation subspace $P_n$ are given by
\ba
P_0&=&(1-n_\up)(1-n_\dn),\\
P_1&=&((1-n_\up)n_\dn + (1-n_\dn)n_\up,\\
P_2&=&n_\up n_\dn.
\ea
The effective Hamiltonian can be written as
\be
H_{\text{eff}}=H_{11}+\sum_{n=0,2}H_{1n}\frac{1}{E-H_{nn}}H_{n1},
\ee
where
\be
H_{mn}=P_mHP_n.
\ee
After some algebra, we find that
\ba
H_{11}&=&H_{NL}\\
H_{01}&=&\sum_\sigma(t\psi^\dagger_\sigma+i\lambda_\sigma\gamma_\sigma)d_\sigma(1-n_{-\sigma}),\\
H_{10}&=&\sum_\sigma(-t\psi_\sigma+i\lambda_\sigma\gamma_\sigma)d^\dagger_\sigma(1-n_{-\sigma}),\\
H_{12}&=&\sum_\sigma(t\psi^\dagger_\sigma+i\lambda_\sigma\gamma_\sigma)d_\sigma n_{-\sigma},\\
H_{21}&=&\sum_\sigma(-t\psi_\sigma+i\lambda_\sigma\gamma_\sigma)d^\dagger_\sigma n_{-\sigma}.
\ea
Using the above expressions with the low energy assumption $E\ll \min(-\epsilon,U-\epsilon)$, we get second order in $t$ and $\lambda$ corrections to the Hamiltonian:
\ba
H_{12}\frac{1}{E-H_{22}}H_{21}&=&\frac{1}{|\epsilon|-U}\sum_{\sigma,\sigma'}\big[ t^2 \psi^\dagger_\sigma \psi_{\sigma'} + \lambda_\sigma\lambda_{\sigma'} \gamma_\sigma \gamma_{\sigma'} + i\lambda_\sigma t\gamma_\sigma\psi_{\sigma'}+i\lambda_{\sigma'}t\gamma_{\sigma'}\psi^\dagger_\sigma \big]d_\sigma n_{-\sigma}d^\dagger_{\sigma'}n_{-\sigma'}\nonumber\\
&=&\sum_{\sigma,\sigma'}D_{\sigma,\sigma'},\label{eq:1221}\\
H_{10}\frac{1}{E-H_{00}}H_{01}&=&-\frac{1}{|\epsilon|}\sum_{\sigma,\sigma'}\big[ t^2 \psi_\sigma \psi^\dagger_{\sigma'} + \lambda_\sigma\lambda_{\sigma'}  \gamma_\sigma \gamma_{\sigma'} - i\lambda_\sigma t\gamma_\sigma\psi^\dagger_{\sigma'}-i\lambda_{\sigma'}t\gamma_{\sigma'}\psi_\sigma \big]d^\dagger_\sigma \bar{n}_{-\sigma}d_{\sigma'}\bar{n}_{-\sigma'}\nonumber\\
&=&\sum_{\sigma,\sigma'}Z_{\sigma,\sigma'}\label{eq:1001},
\ea
where $\bar{n}_\sigma=1-n_{\sigma}$, and $D_{\sigma,\sigma'}$ and $Z_{\sigma,\sigma'}$ are given by
\ba
D_{\up\up}&=&\frac{1}{|\epsilon|-U}\big[t^2\psi^\dagger_\up \psi_\up +\lambda^2 + i\lambda t \gamma_\up(\psi_\up + \psi^\dagger_\up)  \big]n_\dn,  \\
D_{\dn\dn}&=&\frac{1}{|\epsilon|-U}\big[t^2\psi^\dagger_\dn \psi_\dn +\lambda^2 - i\lambda t \gamma_\dn(\psi_\dn + \psi^\dagger_\dn)  \big]n_\up,  \\
D_{\up\dn}&=&\frac{1}{|\epsilon|-U}\big[t^2\psi^\dagger_\up\psi_\dn -\lambda^2\gamma_\up\gamma_\dn + i\lambda t (\gamma_\up \psi_\dn - \gamma_\dn \psi^\dagger_\up)  \big](-d^\dagger_\dn d_\up),  \\
D_{\dn\up}&=&\frac{1}{|\epsilon|-U}\big[t^2\psi^\dagger_\dn\psi_\up -\lambda^2\gamma_\dn\gamma_\up - i\lambda t (\gamma_\dn \psi_\up - \gamma_\up \psi^\dagger_\dn)  \big](-d^\dagger_\up d_\dn), \\
Z_{\up\up}&=&\frac{1}{|\epsilon|}\big[t^2\psi^\dagger_\up \psi_\up - \lambda^2 + i\lambda t \gamma_\up(\psi_\up + \psi^\dagger_\up)  \big]n_\up,  \\
Z_{\dn\dn}&=&\frac{1}{|\epsilon|}\big[t^2\psi^\dagger_\dn \psi_\dn - \lambda^2 - i\lambda t \gamma_\dn(\psi_\dn + \psi^\dagger_\dn)  \big]n_\dn,  \\
Z_{\up\dn}&=&\frac{1}{|\epsilon|}\big[t^2\psi^\dagger_\dn\psi_\up +\lambda^2\gamma_\up\gamma_\dn + i\lambda t (\gamma_\up \psi^\dagger_\dn - \gamma_\dn \psi_\up)  \big]d^\dagger_\up d_\dn,  \\
Z_{\dn\up}&=&\frac{1}{|\epsilon|}\big[t^2\psi^\dagger_\up\psi_\dn +\lambda^2\gamma_\dn\gamma_\up - i\lambda t (\gamma_\dn \psi^\dagger_\up - \gamma_\up \psi_\dn)  \big]d^\dagger_\dn d_\up.
\ea
Introducing the spin operators, $\vec{S}=d^\dagger_\alpha \boldsymbol{\sigma}_{\alpha\beta} d_\beta$, $\vec{s(0)}=\psi^\dagger_\alpha(0) \boldsymbol{\sigma}_{\alpha\beta} \psi_\beta(0)$ and $\vec{S^{\gamma}}=\gamma_\alpha \boldsymbol{\sigma}_{\alpha\beta} \gamma_\beta$ simplifies Eq.(\ref{eq:1221}) and (\ref{eq:1001}):
\ba
H_{12}\frac{1}{E-H_{22}}H_{21}&=&\frac{1}{U-|\epsilon|}\bigg[\frac{t^2}{2}(\vec{S}\cdot \vec{s}(0)-n_\up(0)-n_\dn(0))-\frac{\lambda^2}{2}(S_y S^\gamma_y+2)+\frac{i\lambda t}{2}\bigg\{(\gamma_\up(\psi_\up+\psi^\dagger_\up)+\gamma_\dn(\psi_\dn+\psi^\dagger_\dn))S_z \nonumber\\
&+& \gamma_\up(\psi_\dn S^-+\psi^\dagger_\dn S^+)-\gamma_\dn(\psi_\up S^+ +\psi^\dagger_\up S^-) -(\gamma_\up(\psi_\up+\psi^\dagger_\up)-\gamma_\dn(\psi_\dn+\psi^\dagger_\dn))\bigg\} \bigg],\\
H_{10}\frac{1}{E-H_{00}}H_{01}&=&\frac{1}{|\epsilon|}\bigg[\frac{t^2}{2}(\vec{S}\cdot \vec{s}(0)+n_\up(0)+n_\dn(0))-\frac{\lambda^2}{2}(S_y S^\gamma_y-2)+\frac{i\lambda t}{2}\bigg\{(\gamma_\up(\psi_\up+\psi^\dagger_\up)+\gamma_\dn(\psi_\dn+\psi^\dagger_\dn))S_z \nonumber\\
&+& \gamma_\up(\psi_\dn S^- +\psi^\dagger_\dn S^+)-\gamma_\dn(\psi_\up S^+ +\psi^\dagger_\up S^-) +(\gamma_\up(\psi_\up+\psi^\dagger_\up)-\gamma_\dn(\psi_\dn+\psi^\dagger_\dn))\bigg\} \bigg],
\ea
By ignoring the terms correspond to small shifts in chemical potential, we finally get the low energy effective Hamiltonian $H_b$ in Section
\ref{sec:QDmodel}.

\section{Green's functions in the slave-boson mean-field theory}\label{app:SB1}
To evaluate the correlation functions in Eqs.~\eqref{eq:mfe1} and  \eqref{eq:mfe2}, we first transform the action to the Matsubara frequency domain after the mean-field approximation:
\begin{align}
S_{\text{sb}}=& \sum_{n,\sigma} \bigg[\sum_k \psi^*_{k,n,\sigma}(-i\omega_n+\xi_k)\psi_{k,n,\sigma}+f^*_{n,\sigma}(-i\omega_n+\tilde{\epsilon})f_{n,\sigma} + i\lambda_\sigma b\gamma^1_{-n,\sigma}(f_{n,\sigma} + f^*_{-n,\sigma})+\sum_k tb(f^*_{n,\sigma}\psi_{k,n,\sigma} +\psi^*_{k,n,\sigma}f_{n,\sigma} ) \nonumber\\
&-\frac{1}{2}\sum_{i=1,2}i\omega_n\gamma^i_{-n,\sigma} \gamma^i_{n,\sigma}+i\delta_{1\sigma}\gamma^1_{-n,\sigma}\gamma^2_{n,\sigma}+i\delta_2\gamma^1_{-n,\sigma}\gamma^2_{n,-\sigma} \bigg]\label{eq:Smf},
\end{align}
where $\tilde{\epsilon}=\epsilon+\eta$. Next, we integrate out the NL fermion fields $\psi^*$ and $\psi$ to find the following effective action:
\ba
S_{\text{eff}}(f,\gamma^1,\gamma^2)&=& \sum_{n,\sigma} \bigg[f^*_{n,\sigma}(-i\omega_n+\tilde{\epsilon}+\sum_{k}\frac{t^2b^2}{i\omega_n-\xi_k})f_{n,\sigma} + i\lambda_\sigma b\gamma^1_{-n,\sigma}(f_{n,\sigma} + f^*_{-n,\sigma})-\sum_{i=1,2}\frac{i\omega_n}{2}\gamma^i_{-n,\sigma} \gamma^i_{n,\sigma} \nonumber\\
&&+i\delta_{1\sigma}\gamma^1_{-n,\sigma}\gamma^2_{n,\sigma}+i\delta_2\gamma^1_{-n,\sigma}\gamma^2_{n,-\sigma} \bigg],\\
&=& \sum_{n,\sigma} \bigg[f^*_{n,\sigma}(-i(\omega_n+\Gamma_n)+\tilde{\epsilon})f_{n,\sigma} + i\lambda_\sigma b\gamma^1_{-n,\sigma}(f_{n,\sigma} + f^*_{-n,\sigma})-\sum_{i=1,2}\frac{i\omega_n}{2}\gamma^i_{-n,\sigma} \gamma^i_{n,\sigma}\nonumber\\
&&+i\delta_{1\sigma}\gamma^1_{-n,\sigma}\gamma^2_{n,\sigma}+i\delta_2\gamma^1_{-n,\sigma}\gamma^2_{n,-\sigma} \bigg],
\ea
where $\Gamma_n=\Gamma b^2 \sgn \omega_n$ and $\Gamma=\pi t^2 \nu_F$.
To compute the correlation functions in the mean-field equation, we perform a canonical transformation for the Majorana fields $\gamma^1_\sigma=(c^*_\sigma+c_\sigma)/\sqrt{2}$, $\gamma^2_\up=i(c^*_\up-c_\up)/\sqrt{2}$ and $\gamma^2_\dn=-i(c^*_\dn-c_\dn)/\sqrt{2}$. Then the effective action can be reads
\ba
S_{\text{eff}}(f,c)&=&\sum_{n,\sigma}\bigg[f^*_{n,\sigma}(-i(\omega_n+\Gamma_n)+\tilde{\epsilon})f_{n,\sigma} + \frac{i\lambda_\sigma b}{\sqrt{2}}(c^*_{n,\sigma}f_{n,\sigma}+c_{-n,\sigma}f^*_{-n,\sigma}+c_{-n,\sigma}f_{n,\sigma} + c^*_{n,\sigma}f^*_{-n,\sigma})-i\omega_n c^*_{n,\sigma}c_{n,\sigma}\nonumber\\
&&+\delta_1 c^*_{n,\sigma}c_{n,\sigma}+\delta_2(c^*_{n,\up}c^*_{-n,\dn}-c_{n,\up}c_{-n,\dn})  \bigg]\label{eq:fc}
\ea
 We now introduce the Nambu space and rewrite $S_{\text{eff}}=\sum_{n>0} \phi^\dagger_n A_n \phi_n$ with $\phi^\dagger_n= ( f^*_{n,\up}, \, f^*_{n,\dn}, \,  c^*_{n,\up}, \, c^*_{n,\dn}, \, f_{-n,\up}, \, f_{-n,\dn}, \,  c_{-n,\up}, \, c_{-n,\dn} )$ and
\be
A_n=\left(\!\begin{array}{cccccccc}
\tilde{\epsilon}\!-\!i(\omega_n\!+\!\Gamma_n) & 0 & -\frac{ib\lambda}{\sqrt{2}} & 0 & 0 & 0 & -\frac{ib\lambda}{\sqrt{2}} & 0 \\
0 & \tilde{\epsilon}\!-\!i(\omega_n\!+\!\Gamma_n) & 0 & \frac{ib\lambda}{\sqrt{2}} & 0 & 0 & 0 & \frac{ib\lambda}{\sqrt{2}} \\
\frac{ib\lambda}{\sqrt{2}} & 0 & -\!i\omega_n\!+\!\delta_1 & 0 & i\frac{ib\lambda}{\sqrt{2}} & 0 & 0 & \delta_2 \\
0 & -\frac{ib\lambda}{\sqrt{2}} & 0 & -\!i\omega_n\!+\!\delta_1 & 0 & -i\frac{ib\lambda}{\sqrt{2}} & -\delta_2 & 0 \\
0 & 0 & -i\frac{ib\lambda}{\sqrt{2}} & 0 & -\!\tilde{\epsilon}\!-\! i(\omega_n\!+\!\Gamma_n) & 0 & -\frac{ib\lambda}{\sqrt{2}} & 0 \\
0 & 0 & 0 & i\frac{ib\lambda}{\sqrt{2}} & 0 & -\!\tilde{\epsilon}\!-\! i(\omega_n\!+\!\Gamma_n) & 0 & \frac{ib\lambda}{\sqrt{2}} \\
\frac{ib\lambda}{\sqrt{2}} & 0 & 0 & -\delta_2 & i\frac{ib\lambda}{\sqrt{2}} & 0 & -\!i\omega_n\!-\!\delta_1 & 0 \\
0 & -\frac{ib\lambda}{\sqrt{2}} & \delta_2 & 0 & 0 & -i\frac{ib\lambda}{\sqrt{2}} & 0 & -\!i\omega_n\!-\!\delta_1
\end{array}\!\right).
\ee
The correlation functions can be calculated as
\ba
G_{1}(\omega_n)&\equiv&\langle f_{n,\up} \gamma^1_{-n,\up} \rangle =\frac{1}{\sqrt{2}}(\langle f_{n,\up} c_{-n,\up} \rangle+\langle f_{n,\up}c^*_{n,\up} \rangle) \\
&=&\frac{\Theta(n)}{\sqrt{2}}\left([A_n^{-1}]_{17} +[A_n^{-1}]_{13}\right)-\frac{\Theta(-n)}{\sqrt{2}}\left([A_{-n}^{-1}]_{35} +[A_{-n}^{-1}]_{75}\right)\\
&=& \frac{\omega_n}{i(\omega_n+\Gamma_n)-\tilde{\epsilon}}\cdot\frac{\lambda b}{\omega_n^2+\delta_1^2+\delta_2^2+\frac{2b^2\lambda^2\omega_n(\omega_n+\Gamma_n)}{(\omega_n+\Gamma_n)^2+\tilde{\epsilon}^2}}\\
&=&-\langle  f_{n,\dn}\gamma^1_{-n,\dn} \rangle,\\
G_{f}(\omega_n)&\equiv&\langle f_{n,\up}f^*_{n,\up} \rangle=\Theta(n)[A_n^{-1}]_{11}-\Theta(-n)[A_{-n}^{-1}]_{55}=\frac{-1+i\lambda b G_1(\omega_n)}{i(\omega_n+\Gamma_n)-\tilde{\epsilon}}\\
&=&\langle f_{n,\dn}f^*_{n,\dn} \rangle.
\ea
Notice the following relationship between correlation functions
\be
\langle f^*_{-n,\up}\gamma^1_{-n,\up}\rangle =-\langle f^*_{-n,\dn}\gamma^1_{-n,\dn}\rangle =-G_1(\omega_n)^*.
\ee
To compute $\langle \psi^*_{k,n,\sigma}f_{n,\sigma} \rangle$, we have to integrate out NL fermions $\psi^*_{k',\sigma}$ and $\psi_{k',\sigma}$ for all $k'\neq k$ from Eq.~\ref{eq:Smf}. This procedure leaves the terms $\sum_{n,\sigma}\psi^*_{k,n,\sigma}(-i\omega_n+\xi_k)\psi_{k,n,\sigma}$ and $tb(f^*_{n,\sigma}\psi_{k,n,\sigma} + \psi^*_{k,n,\sigma}f_{n,\sigma})$ in the effective action and shifts $i\Gamma_n\rightarrow i\Gamma_n+\frac{t^2b^2}{i\omega_n-\xi_k}$ such that $S_{\text{eff}}'=\sum_{n>0}\Phi^\dagger B_n \Phi$ where $\Phi^\dagger=(\psi^*_{k,n,\up}, \, \psi^*_{k,n,\dn}, \, \psi_{-n,\up}, \, \psi_{-n,\dn}, \, f^*_{n,\up}, \, f^*_{n,\dn}, \,  c^*_{n,\up}, \, c^*_{n,\dn}, \, f_{-n,\up}, \, f_{-n,\dn}, \,  c_{-n,\up}, \, c_{-n,\dn})$ and
\be
B_n=\left(\begin{array}{cccccccccccc}
\xi_k-i\omega_n & 0 & 0 & 0 & bt & 0 & 0 & 0 & 0 & 0 & 0 & 0 \\
0 & \xi_k-i\omega_n & 0 & 0 & 0 & bt & 0 & 0 & 0 & 0 & 0 & 0 \\
0 & 0 & -\xi_k-i\omega_n & 0 & 0 & 0 & 0 & 0 & -bt & 0 & 0 & 0 \\
0 & 0 & 0 & -\xi_k-i\omega_n & 0 & 0 & 0 & 0 & 0 & -bt & 0 & 0 \\
bt & 0 & 0 & 0 & &&&&&&& \\
0 & bt & 0 & 0 & &&&&&&& \\
0 & 0 & 0 & 0 & &&&&&&& \\
0 & 0 & 0 & 0 & &&&A_n(i\Gamma_n\rightarrow i\Gamma_n+\frac{t^2b^2}{i\omega_n \sgn(n)-\xi_k})\\
0 & 0 & -bt & 0 & &&&&&&& \\
0 & 0 & 0 & -bt & &&&&&&& \\
0 & 0 & 0 & 0 & &&&&&&& \\
0 & 0 & 0 & 0 & &&&&&&&
\end{array}\right).
\ee
Then straightforward calculation gives
\ba
G_T(k,\omega_n)&\equiv&\langle f_{n,\up}\psi^*_{k,n,\up} \rangle=\Theta(n)[B_n^{-1}]_{51}-\Theta(-n)[B_{-n}^{-1}]_{39}=\frac{tb\,G_f(\omega_n)}{i\omega_n-\xi_k}\\
&=&\langle f_{n,\dn}\psi^*_{k,n,\dn} \rangle
\ea

Let's plug above correlation functions back into the mean-field equations (\ref{eq:mfe1}) and (\ref{eq:mfe2}).
\ba
b^2&-&\frac{2}{\beta}\sum_{n}G_f(\omega_n)e^{i\omega_n0^+}=1 \label{eq:mfe3}\\
2b\eta&-&\frac{4t}{\beta}\sum_{k,n}\re[G_T(k,\omega_n)e^{i\omega_n0^+}]-\frac{4\lambda}{\beta}\sum_{n}\re[iG_1(\omega_n)e^{i\omega_n0^+}]=0\label{eq:mfe4}
\ea
First, we evaluate Matsubara sum in Eq.(\ref{eq:mfe3}) using the conventional analytic continuation method with cut along the real frequency axis due to the non-analyticity of $\sgn(\omega_n)=\sgn(\im\omega)$.
\ba
-\frac{2}{\beta}\sum_{n}G_f(\omega_n)&=&\frac{2}{\beta}\sum_n\left[ \frac{1}{i\omega_n -\tilde{\epsilon} +i\Gamma b^2\sgn(\omega_n)}-\frac{i\lambda G_1(\omega_n)}{i\omega_n-\tilde{\epsilon} + i\Gamma b^2\sgn(\omega_n)}\right]\\
&=&\frac{i}{\pi}\oint d\omega n_F(\omega)\left[ \frac{1}{\omega -\tilde{\epsilon} +i\Gamma b^2\sgn(\im\omega)}-\frac{i\lambda G_1(-i\omega)}{\omega-\tilde{\epsilon} + i\Gamma b^2\sgn(\im\omega)}\right]\\
&=&\frac{1}{\pi}\int_{-\infty}^{\infty} d\omega n_F(\omega)\left[\frac{2\Gamma b^2}{(\omega-\tilde{\epsilon})^2+(\Gamma b^2)^2} +F(b,\eta)\right]\\
&\stackrel{T\rightarrow 0}{\approx}&1-\frac{2}{\pi}\arctan \frac{\tilde\epsilon}{\Gamma b^2}+\frac{1}{\pi}\int_{-\infty}^{0} d\omega F(b,\eta),\label{eq:sgf}
\ea
where
\ba
n_F(\omega)&=&\frac{1}{e^{\beta \omega}+1},\\
F(b,\eta)&=&2\re\left[\frac{i\omega\lambda^2 b^2}{(\omega-\tilde{\epsilon}+i\Gamma b^2)^2}\cdot\frac{1}{\omega^2-\delta_1^2-\delta_2^2-\frac{2b^2\lambda^2(\omega^2+i\omega\Gamma b^2)}{(\omega+i\Gamma b^2)^2-\tilde{\epsilon}^2}} \right].
\ea
Let's discuss first the limit of $\lambda \rightarrow 0$. In this case, first two terms in Eq.(\ref{eq:sgf}) combined with Eq.(\ref{eq:mfe3}) lead to the solution $\tilde\epsilon \approx \frac{\pi}{2}\Gamma b^4\sim b^4$. Thus, in this limit $\tilde\epsilon$ is small and one can consider its effect perturbatively. When $\lambda \neq 0$, one can evaluate the integral involving $F(b,\eta)$ in Eq. \eqref{eq:sgf} to find that it vanishes when $\tilde\epsilon=\delta_1=\delta_2=0$. In the limit of $\delta_{1,2}=0$, we calculate the first order correction in $\tilde\epsilon$ finding that Eq. \eqref{eq:mfe3} becomes
\be
b^2-f(r)\frac{\tilde\epsilon}{\Gamma b^2}+O\left(\frac{\tilde\epsilon^2}{(\Gamma b^2)^2}\right)=0,
\ee
where
\be
f(r)=\int_{-\infty}^0 dx \frac{4(r^2x-x^3)}{\pi(1+x^2)(x^4+(1-4r^2)x^2+4r^4)}
\ee
is a dimensionless positive monotonic function of $r=\lambda/\Gamma b$, see Fig. \ref{fig:fr}.
\begin{figure}
	\centering
  \includegraphics[width=3.5in]{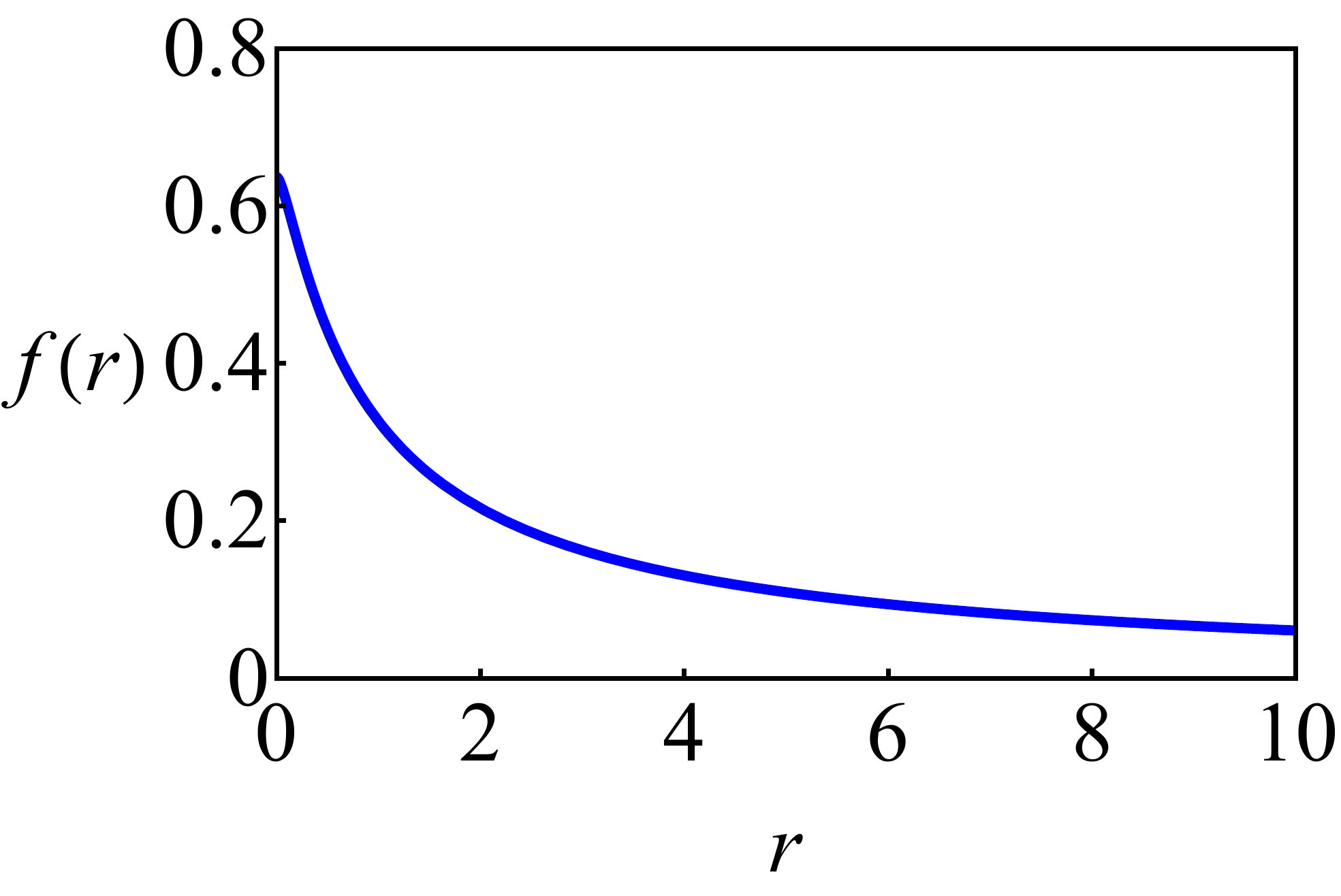}
\caption{The dependence of the dimensionless function $f$ on $r$.}
\label{fig:fr}
\end{figure}
One can see that there exist solution of the above equation even for $\lambda\neq 0$. Indeed, in the limit of small $b$, one can neglect higher order terms in $\tilde\epsilon$, and $\tilde\epsilon\approx \Gamma b^4/f(r)$. For $r=0$, $f(0)=2/\pi$ and for $r\gg 1$, $f(0)\propto 1/r$.

Taking into account the solution for $\tilde\epsilon$, let's evaluate the Matsubara sums in Eq.(\ref{eq:mfe4}):
\ba
-\frac{t}{\beta}\sum_{k,n}G_T(k,\omega_n)&=&\frac{t^2b}{\beta}\sum_{k,n}\frac{1}{i\omega_n-\xi_k}\left[\frac{1}{i\omega_n-\tilde{\epsilon}+i\Gamma b^2\sgn(\omega_n)}-\frac{i\lambda bG_1(\omega_n)}{i\omega_n-\tilde{\epsilon}+i\Gamma b^2\sgn(\omega_n)}\right]\\
&=&-\frac{i\Gamma b}{\beta}\sum_n\left[\frac{\sgn(\omega_n)}{i\omega_n-\tilde{\epsilon}+i\Gamma b^2\sgn(\omega_n)}-\frac{i\lambda bG_1(\omega_n)\sgn(\omega_n)}{i\omega_n-\tilde{\epsilon}+i\Gamma b^2\sgn(\omega_n)}\right]\label{eq:sgt}
\ea
The first term in the above equation can be calculated by introducing a UV cutoff $\Lambda$.
\ba
-\frac{i\Gamma b}{\beta}\sum_n \frac{\sgn(\omega_n)}{i\omega_n-\tilde\epsilon+i\Gamma b^2\sgn(\omega_n)}&=&\frac{\Gamma b}{2\pi} \oint d\omega n_F(\omega) \frac{\sgn(\im\omega)}{\omega-\tilde\epsilon+i\Gamma b^2 \sgn(\im\omega)}\\
&=&\frac{\Gamma b}{\pi} \int_{-\infty}^\infty d\omega n_F(\omega)\frac{\omega-\tilde\epsilon}{(\omega-\tilde\epsilon)^2+(\Gamma b^2)^2} \\
&\stackrel{T\rightarrow 0}{\approx}&\frac{\Gamma b}{\pi} \int_{-\Lambda}^0 d\omega \frac{\omega-\tilde\epsilon}{(\omega-\tilde\epsilon)^2+(\Gamma b^2)^2}\\
&\stackrel{\tilde\epsilon \rightarrow 0}{\approx}&-\frac{\Gamma b}{\pi}\left[\ln \frac{\Lambda}{|\Gamma b^2|}+O\left(\frac{\tilde\epsilon}{\Gamma b^2}\right)\right]
\ea
Since the second term in Eq.(\ref{eq:sgt}) are not UV divergent, we can ignore its contribution. Finally, we proceed to the evaluation of the last term in Eq.(\ref{eq:mfe4}).
\ba
-\frac{\lambda}{\beta}\sum_n iG_1(\omega_n)&=& -\frac{\lambda^2b}{\beta}\sum_n\frac{i\omega_n}{i\omega_n-\tilde\epsilon+i\Gamma b^2\sgn\omega_n} \, \frac{1}{\omega_n^2+\delta_1^2+\delta_2^2+\frac{2b^2\lambda^2\omega_n(\omega_n+\Gamma b^2\sgn\omega_n)}{(\omega_n+\Gamma b^2\sgn\omega_n)^2+\tilde\epsilon^2}}   \\
&=&\frac{1}{\pi b}\int_{-\infty}^{\infty}d\omega n_F(\omega)\re\left[  \frac{i\omega\lambda^2 b^2}{\omega-\tilde{\epsilon}+i\Gamma b^2}\,\frac{1}{\omega^2-\delta_1^2-\delta_2^2-\frac{2b^2\lambda^2(\omega^2+i\omega\Gamma b^2)}{(\omega+i\Gamma b^2)^2-\tilde{\epsilon}^2}}   \right]\\
&\stackrel{T \rightarrow 0}{\approx}&\frac{1}{\pi b}\int_{-\infty}^{0}d\omega\re\left[  \frac{i\omega\lambda^2 b^2}{\omega-\tilde{\epsilon}+i\Gamma b^2}\,\frac{1}{\omega^2-\delta_1^2-\delta_2^2-\frac{2b^2\lambda^2(\omega^2+i\omega\Gamma b^2)}{(\omega+i\Gamma b^2)^2-\tilde{\epsilon}^2}}   \right]\label{eq62}\\
&\stackrel{\delta_1,\delta_2 \rightarrow 0}{\approx}&\frac{\Gamma b}{\pi}\left[\int_{-\infty}^{0}d\omega\frac{\omega\lambda^2 b^2}{(\omega^2-2b^2\lambda^2)^2+\omega^2\Gamma^2b^4}+\mathcal{O}\left(\frac{\tilde\epsilon}{\Gamma b^2}\right)\right]\\
&=&-\frac{\Gamma b}{2\pi}\int_0^\infty dx\frac{1}{x^2+(r^{-2}-4)x+4}+\frac{\Gamma b}{\pi}\mathcal{O}\left(\frac{\tilde\epsilon}{\Gamma b^2}\right)
\ea
where $r=\lambda/\Gamma b$. In the limit of $r\gg 1$ we get,
\be
-\frac{\lambda}{\beta}\sum_n iG_1(\omega_n)\sim-\frac{|\lambda|}{2\sqrt{2}}+\frac{\Gamma b}{\pi}\left[\frac{1}{4}+\mathcal{O}\left(\frac{\tilde\epsilon}{\Gamma b^2}\right)\right]+\mathcal{O}(1/r^2)
\ee
After collecting all the contributions in Eq. (\ref{eq:mfe4}) and neglecting the terms $\mathcal{O}\left(\frac{\tilde\epsilon}{\Gamma b^2}\right)$ and $\mathcal{O}(1/r^2)$, one finds the following equation for $b$:
\be
\eta+\frac{\Gamma}{2\pi}-\frac{2\Gamma}{\pi}\ln\frac{\Lambda}{\Gamma b^2}-\frac{|\lambda|}{\sqrt{2}b}\approx 0.
\ee
Finally, we can ignore the second term since $\eta\sim-\epsilon\gg\Gamma$, and we get Eq. \eqref{eq:sbr}.
\section{Effective action for the gaussian fluctuations in the slave-boson mean-field theory}\label{app:SB2}
Thus far we have neglected effect of quantum fluctuations and considered only mean-field theory. We now take into account Gaussian fluctuations. Around the mean-field saddle point, all the terms linear in fluctuations are absent. The quadradic terms in $\delta s$ and $\dot{\theta}$ can be obtained using Eq. \eqref{eq:effS}.

\begin{align}
S_{\text{eff}}^{(2)}=&-\frac{2}{\beta^2}\sum_{n>0,\nu}\Tr\left[\mathcal{G}_n(\bar{s},\bar{\eta})\delta \mathcal{G}_{2,n,\nu}^{-1}\right]+\frac{1}{\beta^2}\sum_{n>0,\nu}\Tr\left[\mathcal{G}_n(\bar{s},\bar{\eta}) \delta\mathcal{G}_{1,n,-\nu}^{-1} \mathcal{G}_{n+\nu}(\bar{s},\bar{\eta}) \delta\mathcal{G}_{1,n,\nu}^{-1}\right] \\
&+\frac{1}{\beta}\sum_\nu \left[\delta s_{-\nu} (-i\omega_\nu+\bar\eta)\delta s_\nu +2i\bar{s}\delta\dot{\theta}_{-\nu}\delta s_\nu\right]
\end{align}
where the correlation functions are
\ba
\mathcal{G}_{n}(\bar{s}, \bar{\eta})_{1,1}&=&-(G_{f,n}^{-1}\tilde{G}_{f,n}^{-1}\big|_{\bar{s}, \bar{\eta}}-\bar{s}^4G_{\gamma,n}^2)^{-1}\tilde{G}_{f,n}^{-1}\equiv -G_n^p\\
\mathcal{G}_{n}(\bar{s}, \bar{\eta})_{1,2}&=&-(G_{f,n}^{-1}\tilde{G}_{f,n}^{-1}\big|_{\bar{s}, \bar{\eta}}-\bar{s}^4G_{\gamma,n}^2)^{-1}\bar{s}^2G_{\gamma,n}\equiv -\Delta_\sigma\\
\mathcal{G}_{n}(\bar{s}, \bar{\eta})_{2,1}&=&\mathcal{G}_{n}(\bar{s}, \bar{\eta})_{1,2} \equiv -\Delta_n\\
\mathcal{G}_{n}(\bar{s}, \bar{\eta})_{2,2}&=&-(G_{f,n}^{-1}\tilde{G}_{f,n}^{-1}\big|_{\bar{s}, \bar{\eta}}-\bar{s}^4G_{\gamma,n}G_{\gamma,n})^{-1}G_{f,n}^{-1}\equiv -G_n^h,
\ea
\be
\delta\mathcal{G}_{1,n,-\nu}^{-1}=\left(\begin{array}{cc}
i\dot{\theta}_{-\nu}+\delta s_{-\nu}(G_{\psi,n+\nu} +G_{\gamma,n+\nu}+G_{\psi,n} +G_{\gamma,n})\bar{s} & \delta s_{-\nu} (G_{\gamma,n+\nu}+G_{\gamma,n}) \bar{s}\\
\delta s_{-\nu} (G_{\gamma,n+\nu}+G_{\gamma,n}) \bar{s} & -i\dot{\theta}_{-\nu}+\delta s_{-\nu}(\tilde{G}_{\psi,n+\nu} +G_{\gamma,n+\nu}+\tilde{G}_{\psi,n} +G_{\gamma,n})\bar{s}
\end{array}\right),
\ee
\be
\delta\mathcal{G}_{1,n,\nu}^{-1}=\left(\begin{array}{cc}
i\dot{\theta}_{\nu}+\delta s_{\nu}(G_{\psi,n} +G_{\gamma,n}+G_{\psi,n+\nu} +G_{\gamma,n+\nu})\bar{s} & \delta s_{\nu} (G_{\gamma,n+\nu}+G_{\gamma,n}) \bar{s}\\
\delta s_{\nu}(G_{\gamma,n+\nu} + G_{\gamma,n}) \bar{s} & -i\dot{\theta}_{\nu}+\delta s_{\nu}(\tilde{G}_{\psi,n} +G_{\gamma,n}+\tilde{G}_{\psi,n+\nu} +G_{\gamma,n+\nu})\bar{s}
\end{array}\right),
\ee
and
\be
\delta\mathcal{G}_{2,n,\nu}^{-1}=\left(\begin{array}{cc}
\delta s_{-\nu}(G_{\psi,n+\nu}+G_{\gamma,n+\nu})\delta s_{\nu} &  \delta s_{-\nu} G_{\gamma,n+\nu}\delta s_{\nu}\\
\delta s_{-\nu} G_{\gamma,n+\nu}\delta s_{\nu} & \delta s_{-\nu}(\tilde{G}_{\psi,n+\nu}+G_{\gamma,n+\nu})\delta s_{\nu}
\end{array}\right),
\ee
with
\begin{align}
G_{f,n}&=\frac{1}{i\omega_n-\epsilon-\eta-\bar{s}^2(G_{\psi,n}+G_{\gamma,n})}, \\
\tilde{G}_{f,n}&=\frac{1}{i\omega_n+\epsilon+\eta-\bar{s}^2(\tilde{G}_{\psi,n}+G_{\gamma,n})}, \\
G_{\psi,n}&=\tilde{G}_{\psi,n}=-i\Gamma\sgn(n), \quad G_{\gamma,n}=-\frac{i\lambda^2\omega_n}{\omega_n^2+\delta^2}.
\end{align}
After some manipulations, one finds that the fluctuating action reads
\be
S^{(2)}_{\text{eff}}=\frac{1}{2\beta}\sum_\nu (\dot{\theta}_{-\nu} \,\, \delta s_{-\nu}) \left(\begin{array}{cc}
\Gamma^{\dot\theta\dot\theta}_\nu & \bar{s}\Gamma^{\dot\theta s}_\nu \\
\bar{s}\Gamma^{\dot\theta s}_\nu & \Gamma^{ss}_\nu
\end{array}\right) \left(\begin{array}{c}
\dot\theta_\nu \\
\delta s_\nu
\end{array}\right)
\ee
where
\ba
\Gamma^{\dot\theta\dot\theta}_\nu &=&-\frac{2}{\beta}\sum_{n>0}\left[G^p_n G^p_{n+\nu}+G^h_n G^h_{n+\nu}-2\Delta_n\Delta_{n+\nu} \right] \\
\Gamma^{ss}_\nu &=& \Gamma^{s(0)}_\nu+\Gamma^{s(2)}_\nu+\Gamma^{s(4)}_\nu \\
\Gamma^{s(0)}_\nu &=& 2(\bar\eta-i\omega_\nu) \\
\Gamma^{s(2)}_\nu &=& \frac{4}{\beta}\sum_{n>0}\big[G^p_n G_{X,n+\nu}+2\Delta_n G_{\gamma,n+\nu}+G^h_n\tilde{G}_{X,n+\nu}\big] \\
\Gamma^{s(4)}_\nu &=& \frac{4\bar{s}^2}{\beta}\sum_{n>0}\big[G^p_nG_{X,n}G^p_{n+\nu}G_{X,n+\nu} +G_{X,n}G^p_n G_{X,n}G^p_{n+\nu}+G^h_n\tilde{G}_{X,n}G^h_{n+\nu}\tilde{G}_{X,n+\nu} \nonumber \\
&+& \tilde{G}_{X,n} G^h_n \tilde{G}_{X,n} G^h_{n+\nu} + G^p_n G_{\gamma,n} G_{X,n+\nu}\Delta_{n+\nu} + G^p_n G_{\gamma,n} G_{X,n} \Delta_{n+\nu} + G^p_n G_{\gamma,n} G_{X,n+\nu} \Delta_{n}  \nonumber\\
&+& G^p_n G_{\gamma,n+\nu} G_{X,n} \Delta_n + 2G^p_n G_{\gamma,n+\nu} G_{X,n} \Delta_{n+\nu} + G^p_{n+\nu} G_{\gamma,n} G_{X,n} \Delta_n + G^p_n G_{\gamma,n+\nu} G_{X,n+\nu} \Delta_n \nonumber\\
&+& G^h_n G_{\gamma,n+\nu} G_{X,n+\nu} \Delta_n + G^h_n G_{\gamma,n} \tilde{G}_{X,n+\nu} \Delta_{n+\nu} + G^h_n G_{\gamma,n}\tilde{G}_{X,n} \Delta_{n+\nu} + G^h_n G_{\gamma,n}G_{X,n+\nu}\Delta_n\nonumber\\
&+& G^h_n G_{\gamma,n+\nu} G_{X,n} \Delta_n + 2G^h_n G_{\gamma,n+\nu}\tilde{G}_{X,n} \Delta_{n+\nu} + G^h_{n+\nu}G_{\gamma,n} \tilde{G}_{X,n} \Delta_n + 2\Delta_n G_{\gamma,n} \Delta_{n+\nu} G_{\gamma,n+\nu} \nonumber\\
&+& 2\Delta_n G_{\gamma,n}\Delta_{n+\nu} G_{\gamma,n} + 2\Delta_n G_{X,n}\Delta_{n+\nu} \tilde{G}_{X,n+\nu} + 2\Delta_n G_{X,n} \Delta_{n+\nu} \tilde{G}_{X,n}+ 2G^p_n G_{\gamma,n} G^h_{n+\nu} G_{\gamma,n+\nu}\nonumber \\
&+& G^p_n G_{\gamma,n} G^h_{n+\nu} G_{\gamma,n}+ G^p_{n+\nu} G_{\gamma,n} G^h_{n} G_{\gamma,n}\big] \\
\Gamma^{\dot\theta s}_\nu &=& 2i+\frac{2i}{\beta}\sum_{n>0}\big[
G^p_n (G_{X,n}+G_{X,n+\nu}) G^p_{n+\nu} + G^p_n(G_{\gamma,n}+G_{\gamma,n+\nu})\Delta_{n+\nu}   \nonumber\\
&-&G^h_n(\tilde{G}_{X,n}+\tilde{G}_{X,n+\nu})G^h_{n+\nu}-G^h_n(G_{\gamma,n}+G_{\gamma,n+\nu})\Delta_{n+\nu}   \nonumber\\
&+&\Delta_n G_{\gamma,n}(G^p_{n+\nu}-G^h_{n+\nu})+\Delta_n G_{\gamma,n+\nu}(G^p_{n+\nu}-G^h_{n+\nu})   \nonumber\\
&+&\Delta_n (\tilde{G}_{X,n}-G_{X,n})\Delta_{n+\nu}+\Delta_n(\tilde{G}_{X,n+\nu}-G_{X,n+\nu})\Delta_{n+\nu} \big]
\ea
with
\be
G_{X,n}=G_{\psi,n}+G_{\gamma,n}=\tilde{G}_{X,n}.
\ee
Plugging in the mean-field solution $\bar\eta\approx-\epsilon$ leads to
\be
G_{f,n}=\frac{1}{i\omega_n-\bar{s}^2(G_{\psi,n}+G_{\gamma,n})}=\tilde{G}_{f,n} \,\, \rightarrow \,\, G^p_n= G^h_n.
\ee
As a result, we get $\Gamma^{\dot\theta s}_\nu = 2iN$ near the mean-field solution.
\begin{figure}
	\centering
  \includegraphics[width=4.5in]{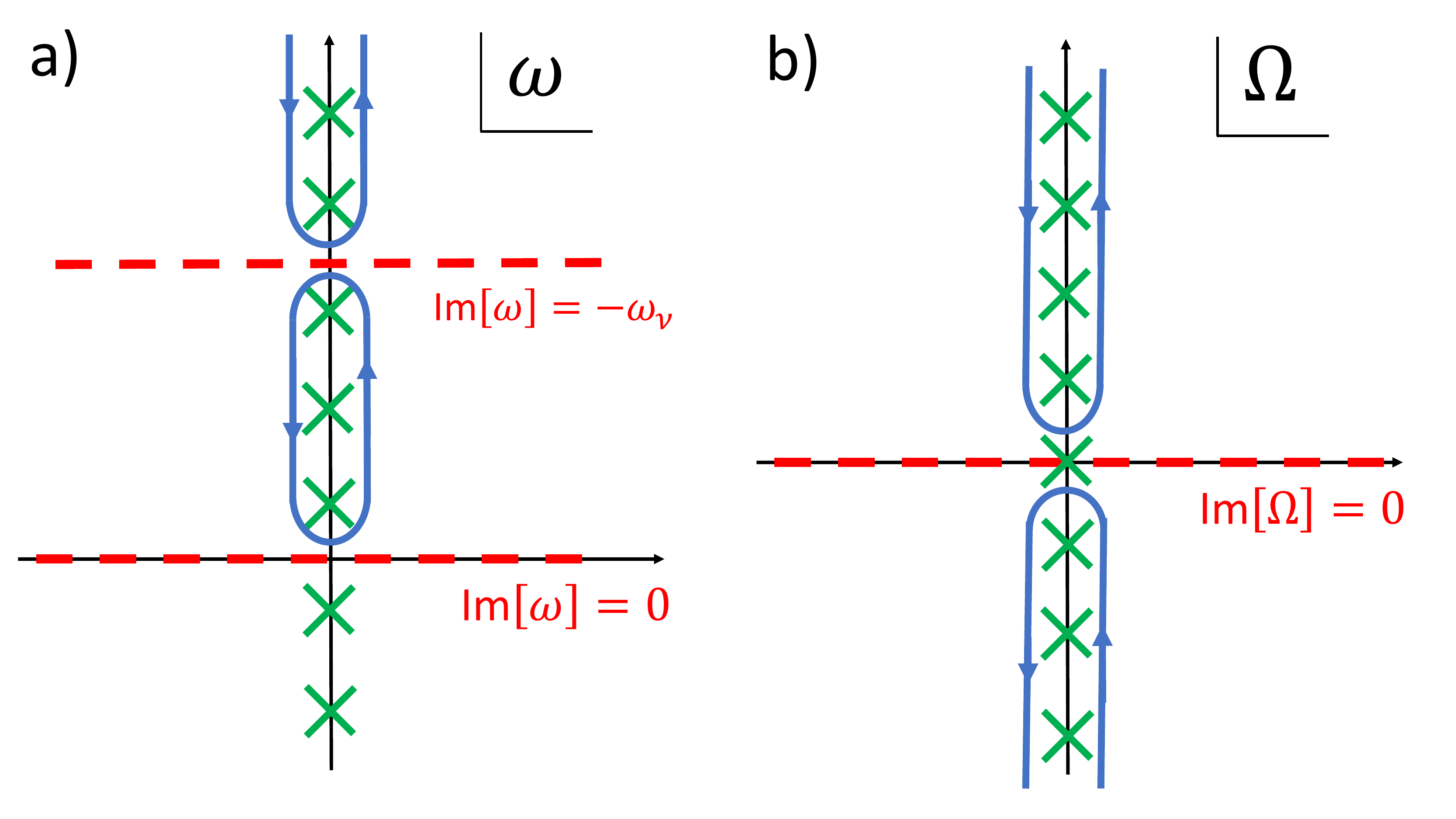}
\caption{a) Integration contour to evaluate fermionic Matsubara sum in $\Gamma_\nu^{ss}$ and $\Gamma_\nu^{\dot\theta\dot\theta}$. b) Integration contour for bosonic Matsubara sum in Eq. \eqref{eq:bcontour}.}
\label{fig:contour}
\end{figure}

To evaluate $\Gamma_\nu^{\dot\theta\dot\theta}$ and $\Gamma_\nu^{ss}$, we need to sum over the fermionic Matsubara frequency $\omega_n$. It can be done using analytical continuation $\omega_n=-i \omega$ and integration along the contour shown in Fig.~\ref{fig:contour} a). One can see that the summation over Matsubara frequency $\omega_n$ can be evaluated by integrating along the branch cuts shown in Fig \ref{fig:contour} a). We note that in addition to the branch cuts, there are also contributions from the poles. However, one can show that the contribution from all the residues sums to zero. Thus, for $\nu>0$ we only have one branch cut $\text{Im}[\omega]=0$. For example,
\ba
\Gamma^{\dot\theta\dot\theta}_{\nu>0} &=& \oint \frac{n_F(\omega)}{2\pi i} \left[G^p(\omega)G^p(\omega+i\omega_\nu)+G^h(\omega)G^h(\omega+i\omega_\nu)-2\Delta(\omega)\Delta(\omega+i\omega_\nu) \right]\nonumber\\
&=&\lim_{\xi \rightarrow 0^+} \int_{-\Lambda}^{\Lambda} \frac{n_F(\omega)}{2\pi i} \left[G^p(\omega+i\xi)G^p(\omega+i\omega_\nu)+G^h(\omega+i\xi)G^h(\omega+i\omega_\nu)-2\Delta(\omega+i\xi)\Delta(\omega+i\omega_\nu) \right]
\ea
 where $n_F(\omega)$ is the Fermi distribution function which we eventually approximate as the theta function in the zero temperature limit; $\Lambda$ is a UV cutoff.
For $\nu<0$ we have two additional integrals above and below branch cut at $\text{Im}[\omega]=\omega_\nu$. Finally, we symmetrize the $\Gamma_\nu^{\dot\theta\dot\theta}$ and $\Gamma_\nu^{ss}$ by averaging the values for $\nu$ and $-\nu$.
Similar method can be used for evaluating the boson correlation function Eq. \eqref{eq:bcontour}. In this case the Matsubara sum can be transformed to an integration over contour shown in Fig. \ref{fig:contour} b).
\end{widetext}
\bibliography{MajoranaKramersRef}

\end{document}